\documentclass[11pt]{article}
\pdfoutput=1

\usepackage[usenames,dvipsnames,svgnames,table]{xcolor} % use color
\usepackage[obeyspaces,hyphens,spaces]{url}
\usepackage{jcapmod}
\usepackage[normalem]{ulem}

\usepackage{shorthand}
\usepackage{mathtools}
\usepackage{booktabs}
\usepackage[english]{babel}
\usepackage{amsmath,amssymb,amsbsy,amstext, amsthm, simplewick, amsfonts,braket}
\usepackage{hyperref}
\usepackage{graphicx}
\usepackage[small]{caption}
\usepackage{slashed}
\usepackage[separate-uncertainty=true,multi-part-units=single]{siunitx}
\usepackage{upgreek}
\usepackage{framed}
\usepackage{wrapfig}
\usepackage{multirow}
\usepackage{bbm}

\usepackage[utf8x]{inputenc}
\usepackage{selinput}
\usepackage{bm}
\usepackage{float}
\usepackage{geometry}
\usepackage{yfonts}
\usepackage{caption}
\usepackage{subcaption}
\usepackage{sidecap}
\usepackage{longtable}
\usepackage{anyfontsize}
\usepackage{dsfont}
\usepackage{tikz}
\usepackage{relsize}
\usepackage{tcolorbox}

\newcommand{\slab}[1]{{\textsc{#1}}}

\newcommand{\floq}[1]{{\scalebox{0.65}{$(#1)$}}}

\usepackage{colortbl}
\definecolor{lightgreen}{cmyk}{0.2, 0, 0.2, 0.2}
\definecolor{lightgray2}{cmyk}{0.1,0.1,0,0.1}
\definecolor{Red2}{RGB}{214, 39, 40}
\definecolor{Blue2}{RGB} {31, 119, 180}
\definecolor{Orange2}{RGB}{255, 127, 14}
\definecolor{Green2}{RGB}{44, 160, 44}
\definecolor{greyish2}{rgb}{.96,.96,.96}

\makeatletter
\newlength{\apb@width}
\newcommand{\autoparbox}[2][c]{\settowidth{\apb@width}{#2}\parbox[#1]{\apb@width}{#2}}

\makeatother


\allowdisplaybreaks[1]
\newcommand{\ped}[1]{\textormath{\textsubscript{#1}}{_{\mathrm{#1}}}}

\DeclarePairedDelimiter{\abs}{\lvert}{\rvert}

\renewcommand{\vec}[1]{\boldsymbol{\mathbf{#1}}}

\newcommand{\dd}{\mathop{\mathrm{d}\!}{}}

\definecolor{pyBlue}{RGB}{31, 119, 180}
\definecolor{pyRed}{RGB}{214, 39, 40}
\definecolor{pyGreen}{RGB}{44, 160, 44}
\definecolor{pyBlue2}{RGB}{0, 111, 237}
\definecolor{pyRed2}{RGB}{224, 52, 36}
\definecolor{Mathematica1}{rgb}{0.368417, 0.506779, 0.709798}
\definecolor{Mathematica2}{rgb}{0.880722, 0.611041, 0.142051}

\def\beq{\begin{equation}}
\def\eeq{\end{equation}}

\begin{document}

\pagenumbering{roman}
\begin{titlepage}
\baselineskip=15.5pt \thispagestyle{empty}

\phantom{h}
\vspace{1cm}
\begin{center}
{\fontsize{21.8}{24}\selectfont  \bfseries  Dynamical Friction in Gravitational Atoms}\\
\end{center}

\vspace{0.5cm}
\begin{center}
{\fontsize{12}{18}\selectfont Giovanni Maria Tomaselli,$^{1}$ Thomas F.M. Spieksma$^{2}$ and Gianfranco Bertone$^{1}$}

\end{center}

\vspace{-0.1cm}
\begin{center}
     \vskip 6pt
\textit{$^1$ Gravitation Astroparticle Physics Amsterdam (GRAPPA), University of Amsterdam,\\
Science Park 904, 1098 XH Amsterdam, The Netherlands}

  \vskip 6pt
\textit{$^2$ Niels Bohr International Academy, Niels Bohr Institute, \\
Blegdamsvej 17, 2100 Copenhagen, Denmark}
\end{center}

\vspace{1.2cm}
\hrule \vspace{0.3cm}
\noindent {\bf Abstract}\\[0.1cm]
Due to superradiant instabilities, clouds of ultralight bosons can spontaneously grow around rotating black holes, creating so-called ``gravitational atoms''. In this work, we study their dynamical effects on binary systems. We first focus on open orbits, showing that the presence of a cloud can increase the cross section for the dynamical capture of a compact object by more than an order of magnitude. We then consider closed orbits and demonstrate that the backreaction of the cloud's ionization on the orbital motion should be identified as dynamical friction. Finally, we study for the first time eccentric and inclined orbits. We find that, while ionization quickly circularizes the binary, it barely affects the inclination angle. These results enable a more realistic description of the dynamics of gravitational atoms in binaries and pave the way for dedicated searches with future gravitational wave detectors.

\vskip10pt
\hrule
\vskip10pt

\end{titlepage}

\thispagestyle{empty}
\setcounter{page}{2}

\tableofcontents

\newpage
\pagenumbering{arabic}
\setcounter{page}{1}

\clearpage

\section{Introduction}

Rotating black holes (BHs) can be used as a probe for fundamental physics. The key behind this is a phenomenon known as \emph{superradiance} \cite{ZelDovich1971,ZelDovich1972,Starobinsky:1973aij,Brito:2015oca}: if an ultralight bosonic field is present in nature, spinning black holes can develop an instability, and a boson cloud can be created around them. Although such bosons have never been detected so far, they can arise in theories beyond the Standard Model, and potentially solve outstanding problems in particle physics and astrophysics. Examples are the QCD axion \cite{Weinberg:1977ma,Wilczek:1977pj,Peccei:1977hh}, axion-like fields from string compactifications \cite{Arvanitaki:2009fg,Svrcek:2006yi} and dark photons \cite{Okun:1982xi,Holdom:1985ag,Cicoli:2011yh}; in this work, we will focus on scalar fields, due to their generally stronger theoretical motivation. Notably, many of these hypothetical particles serve as dark matter candidate \cite{Bergstrom:2009ib,Marsh:2015xka,Hui:2016ltb,Ferreira:2020fam} and are being searched for by several experiments. These searches, however, rely on a non-gravitational coupling of the boson with the Standard Model and, in some cases, on a pre-existing background density of the field.

\vskip 4pt
Neither of these ingredients is needed to trigger the superradiant instability, which can thus be used to probe extremely weakly coupled particles. Rotating BHs naturally shed a significant amount of their energy to the bosonic field: as a result, they spin down and become surrounded by a cloud of ultralight bosons. The BH-cloud system is often called  ``gravitational atom'', due to its structural and mathematical similarity with the hydrogen atom. Such a cloud can manifest its presence in a variety of ways; for example, by emitting a monochromatic gravitational wave (GW) signal which can be picked up by GW detectors. In recent years, another distinctive signature of the cloud has been explored: when a gravitational atom is part of a binary system, a rich phenomenology emerges \cite{Baumann:2018vus, Baumann:2019eav,Baumann:2021fkf,Baumann:2022pkl}. The gravitational waveform from an inspiralling binary could carry direct information about the boson cloud, with invaluable implications for fundamental physics. In this work, we will focus on systems with unequal mass ratios, such as intermediate or extreme-mass ratio inspirals (EMRIs), where the boson cloud is assumed to be around the primary object. This configuration allows one to probe the environment of the central BH optimally with future GW detectors like LISA \cite{LISA:2017pwj,Baker:2019nia} and the Einstein Telescope \cite{Maggiore:2019uih}. With long enough LISA waveforms, it should be possible to identify and interpret waveforms arising from a variety of black hole environments \cite{Cole:2022fir}, and to discriminate EMRIs in presence of gravitational atoms from systems in vacuum, as well as from systems with dark matter overdensities \cite{Gondolo:1999ef,Bertone:2005xz,Eda:2013gg,Eda:2014kra,Yue:2018vtk,Kavanagh:2020cfn,Coogan:2021uqv,Dai:2021olt,Hannuksela:2019vip} and accretion discs \cite{tanaka2002,Derdzinski:2018qzv,Duffell:2019uuk,Derdzinski:2020wlw,Speri:2022upm,Yunes:2011ws,Kocsis:2011dr,Barausse:2014tra}.  

\vskip 4pt
As the companion inspirals around the gravitational atom, it perturbs the cloud with a slowly increasing frequency, which has several consequences. In \cite{Baumann:2018vus, Baumann:2019eav}, it was found that the gravitational perturbation is resonantly enhanced at specific orbital frequencies, around which the cloud is forced to transition (partly or entirely) from one state to another, in analogy with the Landau-Zener process in quantum mechanics. Due to the large shift in energy and angular momentum that such a ``resonant transition'' requires, the backreaction of this process can cause the inspiral to stall or speed up, leaving a distinctive mark on the ensuing waveform. As the binary approaches merger, with its separation becoming comparable to the size of the cloud, the gravitational atom starts to undergo another, different kind of transition: the cloud gets unbound from the parent BH, or \emph{ionized} \cite{Baumann:2021fkf, Baumann:2022pkl}. The energy required by this process is supplied by the binary and it can be overwhelmingly larger than the amount of energy the binary loses by GW emission. As a consequence, the inspiral dynamics is driven, rather than perturbed, by the interaction with the cloud. Although ionization happens at any stage of the inspiral, it features sharp and sudden increments when the orbital frequency raises above certain \emph{thresholds}. These features leave a clear imprint on the gravitational waveform and carry direct information on the gravitational atom, as the position of such thresholds is intimately connected with the boson's mass and the state of the cloud.

\vskip 4pt
In this paper, we aim to investigate the formation and evolution of binary inspirals involving a gravitational atom. First, we study how the presence of a boson cloud affects the binary formation via dynamical capture. This mechanism is well-understood in vacuum, where a soft burst of GWs provides the necessary energy loss to create a bound orbit. When the cloud is present, an additional channel for energy loss opens up, with a consequent increase of the capture cross section and of the binary merger rate. Should the companion indeed be captured, it will be on a very eccentric and, generally, inclined orbit. While previous work on ionization \cite{Baumann:2021fkf, Baumann:2022pkl} assumed quasi-circular and equatorial orbits, we relax these assumptions for the first time. This is not only needed to provide a coherent picture of the binary's evolution, but it is also a necessary step in the direction of a truly realistic description of such systems. The analysis of real GW data will require a fully general understanding of the phenomenology, both for detection and parameter inference. We thus compute how eccentricity and inclination affect ionization, and conversely, how its backreaction affects the orbital parameters. We show that, generally speaking, eccentric orbits circularize under the influence of ionization, while the orbital inclination is barely affected by it. 

\vskip 4pt
All the effects we study in this paper are non-resonant, as both ionization and the energy lost in a dynamical capture do not require the binary to be in a specific configuration. In fact, these interactions can be interpreted as a friction force that continuously acts on the companion throughout its motion. In order to make this point clearest, we show a detailed comparison between ionization and a naive computation of dynamical friction on the companion that moves through the cloud, eventually proving that the two effects should be interpreted as the same. Orbital resonances can be, nevertheless, crucial in the binary's chronological history, as they determine the cloud's state by the time ionization kick in, and they have the potential to stall the inspiral, preventing it to reach merger for an extremely long time. We plan to put together the conclusions of the present paper with the effects of orbital resonances in a future work, with the goal of a complete chronological understanding of these systems.

\paragraph{Outline} The outline of the paper is as follows. In Section~\ref{sec:grav-atom-binaries}, we briefly review superradiance and the spectrum of the gravitational atom, as well as the perturbation induced by the companion. In Section~\ref{sec:capture}, we study how the cloud impacts the formation of binaries via dynamical capture and compute the corresponding capture cross section. In Section~\ref{sec:ionization-dynamical-friction}, we review the ionization of the gravitational atom and describe its interpretation as dynamical friction. In Section~\ref{sec:eccentricity}, we study how eccentricty impacts ionization, and vice versa. In Section~\ref{sec:inclination}, we do a similar exercise, yet now in the case of inclined orbits. Finally, we conclude in Section~\ref{sec:conclusions}. 

\paragraph{Notation and conventions} Throughout this work, we work in natural units ($G=\hbar=c=1$) unless otherwise stated. The central object is assumed to be a Kerr BH, with mass $M$ and dimensionless spin $\tilde a$, with $0\le\tilde a<1$. Its gravitational radius is $r_g=GM/c^2$ and the angular velocity of its horizon is $\Omega_H=\tilde ac/(2r_g(1+\sqrt{1-\tilde a^2}))$. We indicate quantities related to the companion object with an asterisk (e.g.\ $M_*$ is its mass) and those related to the cloud with a lower case ``c''. The mass of the scalar field is denoted by $\mu$, while $\lambda_c=\hbar/\mu c$ is its reduced Compton wavalength. The gravitational fine structure constant is $\alpha=G\mu M/\hbar c$. The cloud is assumed to be mostly in a bound state $\ket{n_b\ell_bm_b}$; we denote other bound states with $\ket{n\ell m}$ and unbound states with $\ket{k;\ell m}$, where $n$, $\ell$ and $m$ are the standard hydrogenic quantum numbers, while $k$ is the continuous wavenumber.

\paragraph{Code availability} The code used in this work and in \cite{Tomaselli:2024bdd} is publicly available on \href{https://github.com/thomasspieksma/GrAB/}{GitHub}.

\section{Gravitational Atoms in Binaries}
\label{sec:grav-atom-binaries}
We start by briefly reviewing the key features of the gravitational atom. In Section~\ref{sec:superradiance}, we describe the superradiance phenomenon and discuss the spectrum of the gravitational atom; then, in Section~\ref{sec:parameters}, we define the parameters of the binary system; finally, in Section \ref{sec:perturbation}, we discuss the gravitational perturbation from the companion.

\subsection{Superradiance and Gravitational Atoms}
\label{sec:superradiance}

Black hole superradiance is a process by which a bosonic field extracts energy and angular momentum from a rotating BH. A bosonic wave, having frequency $\omega$ and azimuthal quantum number $m$ in the BH's frame, is superradiantly amplified if
\beq
\frac\omega{m} < \Omega_H\,,
\label{eqn:superradiance-condition}
\eeq
where $\Omega_H$ is the angular velocity of the event horizon of the BH. Physically, this inequality can be interpreted as the wave being amplified when it has a smaller angular velocity than the BH's horizon it scatters off. Although this process happens both for massive and massless fields, a non-zero mass provides a natural mechanism to trap the waves around the BH. This allows them to continuously undergo superradiant scattering, realizing the ``black hole bomb'' scenario \cite{Press:1972zz,Cardoso:2004nk}, in which the waves are exponentially amplified. In order for the superradiant amplification to be maximally efficient, the two relevant length scales of the problem, the gravitational radius $r_g$ of the BH and the Compton wavelength $\lambda_c$ of the field, need to have roughly the same size:
\beq
\alpha \equiv \frac{r_{g}}{\lambda_{c}} = \mu M \sim  \mathcal{O}(0.01)-\mathcal{O}(1)\,.
\eeq
This ratio, denoted as $\alpha$, is usually referred to as the ``gravitational fine structure constant''.

\vskip 4pt
The equation of motion for a scalar field $\Phi$ with mass $\mu$ in a curved spacetime is the well-known Klein-Gordon equation,
\beq
\Bigl(g^{\alpha\beta}\nabla_\alpha\nabla_\beta-\mu^2\Bigr)\,\Phi(t, \vec{r}) = 0\,,
\label{eqn:KGeqn}
\eeq
where $g_{\alpha\beta}$ is the spacetime metric (in our case, the Kerr metric) and $\nabla_{\alpha}$ is the corresponding covariant derivative. Equation (\ref{eqn:KGeqn}) admits bound state solutions which, in the non-relativistic limit, are very similar to those of the hydrogen atom in quantum mechanics. To show this explicitly, it is convenient to employ the following ansatz:
\beq
\Phi(t, \vec{r})=\frac{1}{\sqrt{2 \mu}}\bigl[\psi(t, \vec{r}) e^{-i \mu t}+\psi^*(t, \vec{r}) e^{+i \mu t}\bigr]\,,
\label{eqn:ansatzwf}
\eeq	
where $\psi(t, \vec{r})$ is a complex scalar field that is assumed to vary on timescales much longer than~$\mu^{-1}$. Substituting (\ref{eqn:ansatzwf}) into (\ref{eqn:KGeqn}), the Klein-Gordon equation reduces, to leading order in $\alpha$, to the Schrödinger equation with a Coulomb-like potential:
\beq
i \frac{\partial}{\partial t} \psi(t, \vec{r})=\biggl[-\frac{1}{2 \mu} \nabla^{2}-\frac{\alpha}{r}\biggr] \psi(t, \vec{r})\,.
\label{eqn:Schrodingereq}
\eeq
By analogy with the hydrogen atom, the eigenstate solutions to the Schrödinger equation can be written as
\beq
\psi_{n \ell m}(t, \vec{r})=R_{n \ell}(r) Y_{\ell m}(\theta, \phi) e^{-i\left(\omega_{n \ell m}-\mu\right) t}\,.
\label{eqn:eigenstates}
\eeq
Here, $n, \ell$ and $m$ are the principal, angular momentum and azimuthal (or magnetic) quantum numbers, respectively, which must obey $n > \ell $, $\ell \geq 0$ and $\ell \geq |m|$; then, $Y_{\ell m}$ are the scalar spherical harmonics and $R_{n\ell}$ the hydrogenic radial functions, defined as
\beq
R_{n \ell}(r)=\sqrt{\left(\frac{2 \mu \alpha}{n}\right)^3 \frac{(n-\ell-1) !}{2 n(n+\ell) !}}\,\biggl(\frac{2 \alpha \mu r}{n}\biggr)^{\ell} \exp \biggl(-\frac{\mu \alpha r}{n}\biggr) L_{n-\ell-1}^{2 \ell+1}\biggl(\frac{2 \mu \alpha r}{n}\biggr)\,,
\label{eqn:HydrogenicRadial}
\eeq
where $L_{n-\ell-1}^{2 \ell+1}(x)$ is the associated Laguerre polynomial. The radial profile of the eigenstate has most of its support around $r \sim n^{2}r\ped{c}$, where $r\ped{c} \equiv(\mu \alpha)^{-1}$ is the Bohr radius, and decays exponentially as $r \rightarrow\infty$.

\vskip 4pt
The analogy with the hydrogen atom, however, is not exact. The main difference arises from the purely ingoing boundary conditions at the BH horizon, which replace the regularity at $r=0$ usually imposed for the hydrogen atom. As a consequence, the eigenstates are generally ``quasi-bound'', with complex eigenfrequencies:
\beq
    \omega_{n \ell m} = (\omega_{n\ell m})_R+i(\omega_{n\ell m})_I\,,
\eeq
where the subscripts $R$ and $I$ denote the real and imaginary parts of $\omega_{n\ell m}$, respectively. Without loss of generality, we can assume $(\omega_{n\ell m})_R>0$. Moreover, it can be shown that modes that satisfy the superradiance condition (\ref{eqn:superradiance-condition}) have $(\omega_{n\ell m})_I>0$: this means that their occupancy number grows exponentially in time, with the consequent formation of a Bose-Einstein condensate around the BH. The process stops when this ``boson cloud'' has extracted enough mass and angular momentum, so that the condition (\ref{eqn:superradiance-condition}) is saturated and $(\omega_{n\ell m})_I=0$. This cloud-BH system is often referred to as a ``gravitational atom''. The fastest-growing state is $\ket{n\ell m}=\ket{211}$, and the maximum possible mass of the cloud is about $0.1M$ \cite{Hui:2022sri,Herdeiro:2021znw,East:2017ovw}.

\vskip 4pt
For $\alpha\ll1$, the energy of the quasi-bound states is given by \cite{Baumann:2019eav}
\beq
\epsilon_{n\ell m}\equiv(\omega_{n\ell m})_R=\mu\left(1-\frac{\alpha^{2}}{2 n^{2}}-\frac{\alpha^{4}}{8 n^{4}}-\frac{(3 n-2 \ell-1) \alpha^{4}}{n^{4}(\ell+1 / 2)}+\frac{2 \tilde{a} m \alpha^{5}}{n^{3} \ell(\ell+1 / 2)(\ell+1)}+\mathcal{O}\left(\alpha^{6}\right)\right)\,.
\label{eq:eigenenergy}
\eeq
Borrowing the terminology from atomic physics, two states are said to have a \emph{Bohr} ($\Delta n\ne0$), \emph{Fine} ($\Delta n=0$, $\Delta\ell\ne0$) or \emph{Hyperfine} ($\Delta n=0$, $\Delta\ell=0$, $\Delta m\ne0$) splitting, depending on the leading power of $\alpha$ remaining when taking the difference of their energies.

\vskip 4pt
The full spectrum of the gravitational atom contains another class of solutions to the Schrödinger equation (\ref{eqn:Schrodingereq}), namely the unbound states. Similar to the bound states, they take on the following form:
\beq
\psi_{k; \ell m}(t, \vec{r})=R_{k; \ell}(r) Y_{\ell m}(\theta, \phi) e^{-i\epsilon(k)t}\,,
\label{eq:unboundstates}
\eeq
where the discrete quantum number $n$ has been replaced by the continuous wavenumber $k$. Here, the radial function is given by
\beq
R_{k ; \ell}(r)=\frac{2 k e^{\frac{\pi \mu \alpha}{2 k}}\bigl|\Gamma\bigl(\ell+1+\frac{i \mu \alpha}{k}\bigr)\bigr|}{(2 \ell+1) !}(2 k r)^{\ell} e^{-i k r}\,{ }_{1}F_{1}\biggl(\ell+1+\frac{i \mu \alpha}{k} ; 2 \ell+2 ; 2 i k r\biggr)\,,
\label{eq:radialcontinuum}
\eeq
where ${ }_{1} F_{1}(a ; b ; z)$ is the Kummer confluent hypergeometric function. Unlike for the bound states, the eigenfrequencies are now real, $\omega(k) = \sqrt{\mu^{2}+k^{2}}$, and the dispersion relation is
\beq
\epsilon(k) = \sqrt{\mu^{2}+k^{2}} - \mu \approx \frac{k^{2}}{2\mu}\,,
\label{eq:dispersionrelation}
\eeq
where the last approximation is only valid in the non-relativistic regime ($k \ll \mu$) we work in.

\subsection{Binary System}
\label{sec:parameters}

\begin{figure}
\centering
\includegraphics[width=0.6\textwidth]{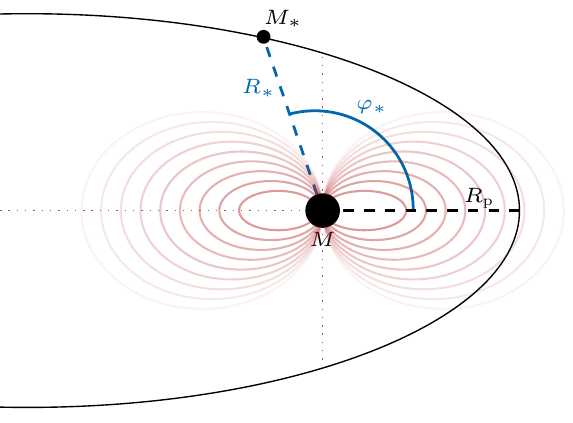}
\caption{Schematic illustration of the binary system we study in this work. The primary object of the binary has mass $M$, while the companion has mass $M_*$. The motion of the companion on the equatorial plane is described by $\{R_{*}, \varphi_*\}$ and $R\ped{p}$ is the periapsis. The red lines schematically indicate the boson~cloud.}
\label{fig:IllustrationParabolic}
\end{figure}

In this work, we deal with generically inclined or eccentric orbits. It is therefore useful to clearly outline the binary system we study. Since we will not deal with inclination until Section~\ref{sec:inclination}, we ignore it here to avoid unnecessary complications. In Figure~\ref{fig:IllustrationParabolic}, we show a schematic illustration of our setup, including the relevant parameters.

\vskip 4pt
We consider a binary system, where the primary object with mass $M$ is much heavier than its companion with mass $M_{*}$, such that the \textit{mass ratio} $q \equiv M_*/M \ll1$. We work in the reference frame of the central BH, where $\vec{r} = \{r, \theta, \phi \}$. The coordinates of the companion are $\vec{R}_{*} = \{R_{*},\theta_*, \varphi_{*} \}$, where $R_{*}$ is the binary's separation and $\theta_*$ is the polar angle with respect to the BH's spin. Since we postpone the discussion of inclined orbits to Section~\ref{sec:inclination}, the orbit entirely lies in the equatorial plane; consequently, we have $\theta_*=\pi/2$, while $\varphi_*$ coincides with the $\emph{true anomaly}$. On a non-circular orbit we denote with $R\ped{p}$ the $\emph{periapsis}$, which is the distance of closest approach between the two components of the binary.

\vskip 4pt
Due to the emission of gravitational waves, the binary inspirals. Consequently, the instantaneous orbital frequency $\Omega(t)$ slowly increases in time. On a Keplerian orbit, the average power and torque emitted by GWs over one period are \cite{Peters:1963ux,Peters:1964zz}
\begin{align}
\label{eq:p_gw}
P_\slab{gw} &= \frac{32}{5}\frac{q^{2}M^{5}(1+q)}{a^{5}(1-\varepsilon^{2})^{7/2}}\biggl(1 +\frac{73}{24}\varepsilon^{2}+\frac{37}{96}\varepsilon^{4}\biggr)\,,\\
\label{eq:tau_gw}
\tau_\slab{gw}&=\frac{32}{5}\frac{q^{2}M^{9/2}\sqrt{1+q}}{a^{7/2}(1-\varepsilon^2)^2}\biggl(1+\frac78\varepsilon^2\biggr)\,,
\end{align}
where we denoted by $a$ the semi-major axis and by $\varepsilon$ the eccentricity of the orbit.

\subsection{Gravitational Perturbation}
\label{sec:perturbation}

The companion object interacts with the cloud gravitationally, introducing a perturbation $V_*$ to the right-hand side of the Schrödinger equation (\ref{eqn:Schrodingereq}). We can write it using the multipole expansion of the Newtonian potential as\footnote{Our convention on the sign of $m_*$ differs from, for example, \cite{Baumann:2021fkf}.}
\beq
\label{eqn:V_star}
V_*(t,\vec r)=-\sum_{\ell_*=0}^\infty\sum_{m_*=-\ell_*}^{\ell_*}\frac{4\pi q\alpha}{2\ell_*+1}Y_{\ell_*m_*}(\theta_*,\varphi_*)Y_{\ell_*m_*}^*(\theta,\phi)\,F(r)\,,
\eeq
where
\beq
F(r)=
\begin{cases}
\dfrac{r^{\ell_*}}{R_*^{\ell_*+1}}\Theta(R_*-r)+\dfrac{R_*^{\ell_*}}{r^{\ell_*+1}}\Theta(r-R_*)&\text{for }\ell_*\ne1\,,\\[12pt]
\biggl(\dfrac{R_*}{r^2}-\dfrac{r}{R_*^2}\biggr)\Theta(r-R_*)&\text{for }\ell_*=1\,,
\end{cases}
\eeq
and $\Theta$ is the Heaviside step function.\footnote{\fontdimen2\font=0.75ex The term with $\ell_*=1$ was incorrectly set to zero in \cite{Baumann:2021fkf, Baumann:2022pkl}. We thank Rodrigo Vicente for pointing this out~\cite{wip_with_Rodrigo}.}

\vskip 4pt
The perturbation induces a mixing between the cloud's bound state $\ket{n_b \ell_b m_b}$ and another state $\ket{n \ell m}$, with matrix element
\beq
\braket{n\ell m|V_*(t,\vec r)|n_b\ell_bm_b}=-\sum_{\ell_*,m_*}\frac{4\pi\alpha q}{2\ell_*+1}Y_{\ell_*m_*}(\theta_*,\varphi_*)\, I_r \, I_{\Omega}\,,
\label{eqn:MatrixElement}
\eeq
where the \emph{radial} and \emph{angular} integrals are
\begin{align}
I_r&=\int_0^\infty F(r)R_{n\ell}(r) R_{n_b \ell_b}(r)\, r^2\dd r\,,\\
\label{eqn:I_Omega}
\begin{split}
I_{\Omega}&= \int Y_{\ell m}^*(\theta, \phi) Y_{\ell_* m_*}^*(\theta, \phi) Y_{\ell_bm_b}(\theta, \phi) \dd \Omega\\
&=\sqrt{\frac{(2 \ell+1)(2 \ell_*+1)(2 \ell_b+1)}{4 \pi}}\begin{pmatrix}
\ell& \ell_* & \ell_b \\
0 & 0 & 0
\end{pmatrix}\begin{pmatrix}
\ell& \ell_* & \ell_b \\
-m & -m_* & m_b
\end{pmatrix}\,.
\end{split}
\end{align}
The expression of $I_\Omega$ in terms of the Wigner-3j symbols implies the existence of \emph{selection rules} that need to be satisfied in order for (\ref{eqn:I_Omega}) to be non-zero:
\begin{align}
(\text{S1})\qquad & {-m}-m_*+m_b=0\,,\\
(\text{S2})\qquad & \ell+\ell_*+\ell_b=2 p,\qquad \text{for }p \in \mathbb{Z}\,,\\
(\text{S3})\qquad & \abs{\ell_b-\ell} \leq \ell_* \leq \ell_b+\ell\,.
\end{align}
Due to the (quasi)-periodicity of $\varphi_{*}(t)$, the matrix element (\ref{eqn:MatrixElement}) can be decomposed into Fourier components as
\beq
\braket{n\ell m|V_{*}(t,\vec r)|n_b \ell_b m_b} = \sum_{g\in \mathbb{Z}} \eta^\floq{g} e^{-i g \Omega t}\,.
\label{eqn:def-eta}
\eeq
To make the notation clearest, we will often remove or add superscripts and subscripts to $\eta^\floq{g}$. Analogous formulae hold for the mixing of $\ket{n_b\ell_bm_b}$ with an unbound state $\ket{k;\ell m}$.

\section{Dynamical Capture}
\label{sec:capture}

The formation of compact binaries is an active area of research (see e.g. \cite{Amaro-Seoane:2012lgq,Amaro-Seoane:2022rxf,Babak:2017tow} and references therein). One of the proposed mechanisms, \emph{dynamical capture}, allows the creation of a bound system through dissipation of energy in a burst of GWs during a close encounter between the two objects. The cross section for this process is \cite{1989ApJ...343..725Q,Mouri:2002mc}
\beq
\sigma_\slab{gw}=2\pi M^2\biggl(\frac{85\pi}{6\sqrt2}\biggr)^{2/7}q^{2/7}(1+q)^{10/7}v^{-18/7}\,,
\label{eqn:gw-cross-section}
\eeq
where the two compact objects have masses $M$ and $M_*=qM$, and $v$ is their relative asymptotic velocity before the close encounter.

\vskip 4pt
When one of the two objects is surrounded by a scalar cloud, then the energy during a dynamical capture is not only emitted via GWs, but also exchanged with the cloud. This phenomenon was first computed in \cite{Zhang:2019eid} and is akin to the ``tidal capture'' found in \cite{Cardoso:2022vpj}. In this section, we will review the computation of the energy exchanged with the bound states of the cloud\footnote{We thank the authors of \cite{Zhang:2019eid} for acknowledging in a private communication the discrepancy with their results.} and extend it to include unbound states as well. Then, we will show how the formula (\ref{eqn:gw-cross-section}) for the cross section gets corrected, and discuss the impact for the merger rate in astrophysically realistic environments.

\subsection{Energy Lost to the Cloud}
\label{sec:E_lost}

In the same spirit as in the derivation of (\ref{eqn:gw-cross-section}), we consider a binary on a parabolic orbit.\footnote{For a non-zero $v$, the orbit is actually hyperbolic. However, approximating it with a parabola allows to correctly compute the leading order in $v$ of the cross section, while greatly simplifying the calculation.} The separation $R_*$ and azimuthal angle $\varphi_*$ can be parametrized as
\beq
R_*=R\ped{p}\biggl(\xi^2-1+\frac1{\xi^2}\biggr)\qquad\text{and}\qquad\varphi_*=2\arctan\biggl(\xi-\frac1\xi\biggr)\,,
\label{eqn:parabola}
\eeq
where $R\ped{p}$ is the periapsis of the orbit and
\beq
\xi\equiv\Bigl(\Omega\ped{p}t+\sqrt{1+\Omega\ped{p}^2t^2}\Bigr)^{1/3}\,,\qquad\text{with}\qquad\Omega\ped{p}\equiv\frac32\sqrt{\frac{(1+q)M}{2R\ped{p}}}\,.
\eeq
Under the gravitational perturbation of the binary, the cloud's wavefunction will evolve with time. It is useful to decompose it as
\beq
\ket{\psi(t)}=\sum_{n,\ell, m}c_{n\ell m}(t)\ket{n\ell m}+\int\frac{\dd k}{2\pi}\sum_{\ell, m}c_{k;\ell m}(t)\ket{k;\ell m}\,.
\eeq
As long as the perturbation is weak enough to keep $\abs{c_{n_b\ell_bm_b}}\approx1$ throughout the evolution, with all other coefficients remaining much smaller, the Schrödinger equation can be approximated as
\beq
i\frac{\dd c_{n\ell m}}{\dd t}\approx \braket{n\ell m|V_*(t,\vec r)|n_b\ell_bm_b}e^{i(\epsilon_{n\ell m}-\epsilon_b)t}\,,
\label{eqn:schrodinger-approx}
\eeq
where $\epsilon_b$ is the energy of $\ket{n_b\ell_bm_b}$. In the limit $t\to+\infty$, equation (\ref{eqn:schrodinger-approx}) can then be integrated to give
\beq
c_{n\ell m}=-i\int_{-\infty}^{+\infty}\dd t\braket{n\ell m|V_*(t,\vec r)|n_b\ell_bm_b}e^{i(\epsilon_{n\ell m}-\epsilon_b)t}\,.
\label{eqn:cnlm-b}
\eeq
An identical formula holds for unbound states, where the principal quantum number $n$ is replaced by the continuous wavenumber $k$.

\vskip 4pt
The coefficients $c_{n\ell m}$ and, especially, $c_{k;\ell m}$ are computationally expensive to determine, as they feature the radial integral $I_r$ nested inside the time integral appearing in (\ref{eqn:cnlm-b}). Restricting to an equatorial orbit, where $\theta_*=\pi/2$, they can be written as
\beq
c_{n\ell m}=-i\sum_{\ell_*}\frac{4\pi\alpha q}{2\ell_*+1}I_\Omega Y_{\ell_*g}\biggl(\frac\pi2,0\biggr)\times2\int_0^\infty\dd t\,I_r(t)\cos(g\varphi_*+(\epsilon_{n\ell m}-\epsilon_b)t)\,,
\eeq
and similarly for $c_{k;\ell m}$, where $g=\pm(m-m_b)$ for co/counter-rotating orbits, respectively. The radial integral $I_r$ depends on the time $t$ through $R_*$, as determined in (\ref{eqn:parabola}). Once $c_{n\ell m}$ and $c_{k;\ell m}$ are known, the total energy lost by the binary to the cloud is then given by
\beq
E\ped{lost}=\frac{M\ped{c}}\mu\sum_{n,\ell, m}(\epsilon_{n\ell m}-\epsilon_b)\abs{c_{n\ell m}}^2+\frac{M\ped{c}}\mu\int\frac{\dd k}{2\pi}\sum_{\ell, m}(\epsilon(k)-\epsilon_b)\abs{c_{k;\ell m}}^2\,.
\label{eqn:E_lost}
\eeq
Note that the contribution due to bound states can in principle be negative, due to the existence of states with lower energy, while the term associated to unbound states can only be positive.

\vskip 4pt
Equation (\ref{eqn:E_lost}) requires a sum over an infinite number of final states. For the first term, the one corresponding to transitions to other bound states, we truncate the sum when the addition of a term with higher $n$ would change the result by less than $0.1\%$. This typically requires including terms up to $n\sim10$ to $n\sim35$, depending on the chosen value of $R\ped{p}$. The second term is harder to handle, as there is yet another integral, over the wavenumber $k$. Moreover, for a fixed $k$, all values of $\ell$ are allowed. We evaluate the integrand at discrete steps in $k$, truncating the sum over $\ell$ when the addition of a new term would change the result by less than $0.01\%$. The size of the step depends on the value of $R\ped{p}$ and is chosen to be small enough to properly sample the integrand. The integral over $\dd k$ is then performed with a simple trapezoidal approximation.

\begin{figure}
\centering
\includegraphics{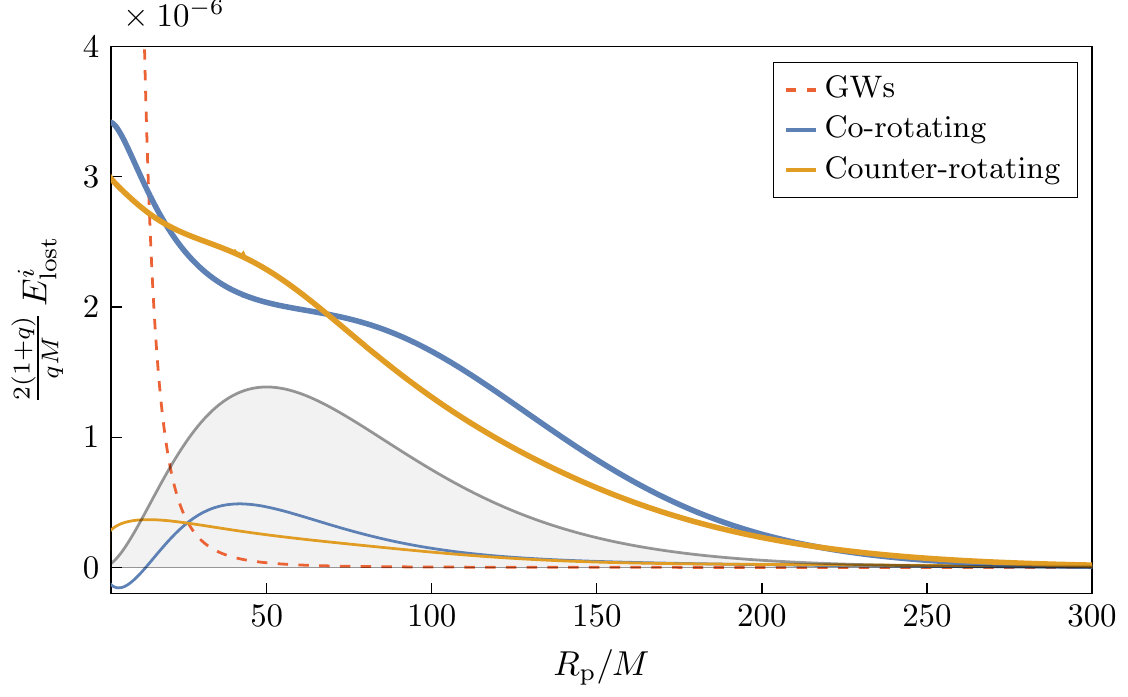}
\caption{Energy lost to the cloud (\ref{eqn:E_lost}) as function of the distance of closest approach $R\ped{p}$, for $\alpha=0.2$, $q=10^{-3}$, $M\ped{c}=0.01M$ and a cloud in the $\ket{211}$ state. The $i$ subscript denotes each of the two contributions to (\ref{eqn:E_lost}), so that $E\ped{lost} = \sum_iE\ped{lost}^{i}$. Thick (thin) lines refer to the energy lost to unbound (bound) states, while the colors differentiate between co-rotating and counter-rotating orbits. For comparison, we also show the density profile of the cloud, $\abs{\psi(R\ped{p})}^2$, in shaded gray, arbitrarily normalized.}
\label{fig:E_lost_211}
\end{figure}

\vskip 4pt
The results are shown in Figure~\ref{fig:E_lost_211}. Here, we plot $E\ped{lost}$, normalized by\footnote{The motivation behind this normalization will become clear in equation (\ref{eqn:E_lost-Kin}).} $qM/(2(1+q))$, for the state $\ket{211}$ and a fiducial set of parameters. Both the contribution due to bound states and to unbound states vanish exponentially for $R\ped{p}\to\infty$ and are largest when $R\ped{p}$ is roughly comparable to the size of the cloud. We also see that the dominant contribution to $E\ped{lost}$ is the one associated to unbound states. At very small radii, $E\ped{lost}$ has a finite limit, meaning that the cloud is only able to dissipate a certain maximum amount of energy. On the other hand, $E_\slab{gw}$ (i.e.\ the energy radiated in GWs) formally diverges for $R\ped{p}\to0$, implying that the high-$v$ limit is dominated by GWs, which become much more effective than the cloud at dissipating energy. Because $E_\slab{gw}$ decays polynomially for $R\ped{p}\to\infty$, GWs will also dominate the low-$v$ limit.

\subsection{Scalings and Cross Section}
\label{sec:cross-section}

Although the values presented in Figure~\ref{fig:E_lost_211} are computed for a fiducial set of parameters, in the limit of small $q$ an approximate scaling relation allows us to predict the values for an arbitrary set of parameters. In the same spirit as equations (3.31) and (3.32) of \cite{Baumann:2021fkf}, we can exploit the $\alpha$-scaling of the radial wavefunctions and of the overlap integrals to write
\beq
E\ped{lost}=M\ped{c}\alpha^2q^2\mathcal E(\alpha^2R\ped{p}/M)\,,
\label{eqn:E_lost-scaling}
\eeq
where $\mathcal E$ is a function that only depends on the initial state $\ket{n_b\ell_bm_b}$.

\vskip 4pt
Once $E\ped{lost}$ is known, we can use it to determine the total cross section $\sigma\ped{tot}$ for dynamical capture by requiring that it is larger than the total initial energy of the binary:
\beq
E\ped{lost}+E_\slab{gw}>\frac12\frac{qM}{1+q}v^2\,,
\label{eqn:E_lost-Kin}
\eeq
where we took into account the contribution due to GW emission,
\beq
E_\slab{gw}=\frac{85\sqrt2}{24}\frac{q^2M^{9/2}}{R\ped{p}^{7/2}}\,.
\eeq
If the left-hand side of (\ref{eqn:E_lost-Kin}) were a decreasing function of $R\ped{p}$, the inequality would hold for all $R\ped{p}<R\ped{p}(v)$ for some function $R\ped{p}(v)$. By relating this to the binary's impact parameter $b$, we would find the total cross section as
\beq
\sigma\ped{tot}=\pi b^2,\qquad\text{where}\qquad b^2=\frac{2MR\ped{p}}{v^2}\,.
\eeq
In reality, while $E_\slab{gw}$ is indeed a decreasing function, $E\ped{lost}$ in general is not. The inequality (\ref{eqn:E_lost-Kin}) will then hold in some finite intervals of $R\ped{p}$. Consequently, for some values of $v$, the cross section for dynamical capture should be geometrically interpreted as an annulus (or several concentrical annuli), rather than a circle.

\begin{figure}
\centering
\includegraphics{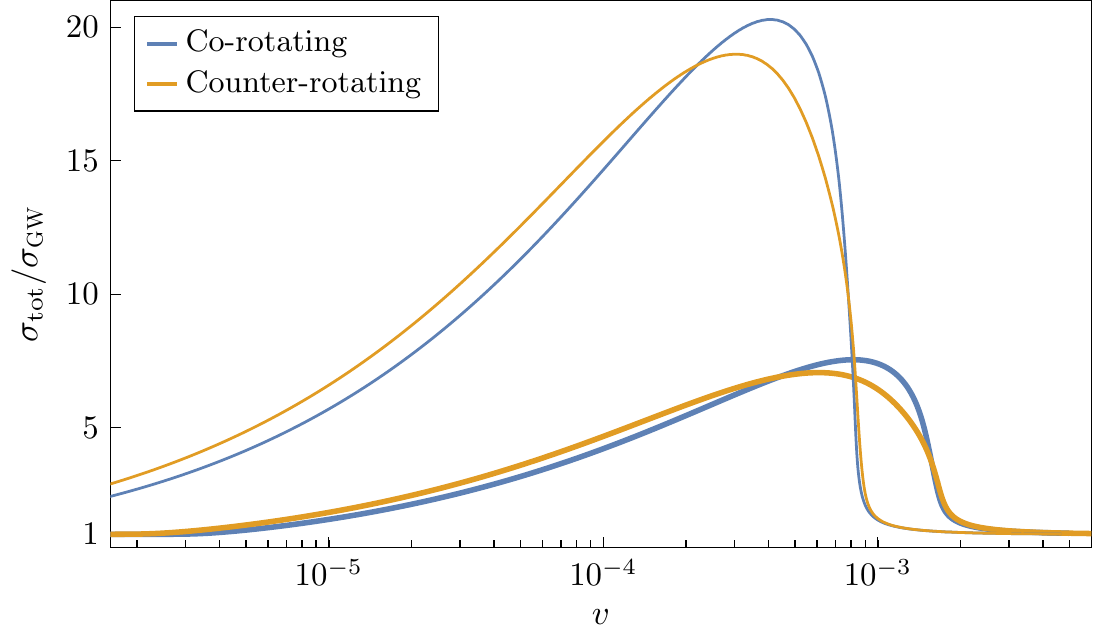}
\caption{Capture cross section $\sigma\ped{tot}$, including the energy lost to both the cloud and GWs, normalized by capture cross section (\ref{eqn:gw-cross-section}) due to GWs only. The cross section is shown as a function of the relative asymptotic velocity between the two objects, $v$. Thick lines are computed for the same set of parameters as in Figure \ref{fig:E_lost_211}, while thin lines show the result when $\alpha$ is decreased from $0.2$ to $0.1$.}
\label{fig:capture_cross_section}
\end{figure}

\vskip 4pt
The results are shown in Figure~\ref{fig:capture_cross_section}. As anticipated in Section~\ref{sec:E_lost}, the ratio $\sigma\ped{tot}/\sigma_\slab{gw}$ asymptotes to unity for very high and very low values of $v$. For intermediate velocities, instead, the cross section is significantly enhanced by the presence of the cloud, which dominates over GWs. The magnitude of the enhancement and the velocities at which it occurs depend on the chosen parameters. In general, the total cross section $\sigma\ped{tot}$ does not inherit any scaling relation akin to (\ref{eqn:E_lost-scaling}), because $E_\slab{gw}$ and $E\ped{lost}$ scale differently with the parameters. However, in the region of parameter space where $E\ped{lost}\gg E_\slab{gw}$, we can neglect the latter in (\ref{eqn:E_lost-Kin}) and derive an approximate scaling relation that reads
\beq
\frac{\sigma\ped{tot}}{\sigma_\slab{gw}}\sim\biggl(\frac{M\ped{c}}M\biggr)^{2/7}\alpha^{-10/7}\,\mathcal S\biggl(\frac{Mv^2}{M\ped{c}\alpha^2q}\biggr)\,,
\label{eqn:scaling-cross-section}
\eeq
where again $\mathcal S$ is a universal function to be found numerically. Equation (\ref{eqn:scaling-cross-section}) allows us to rescale the results of Figure~\ref{fig:capture_cross_section} for other values of the parameters. In particular, it shows that, for smaller values of $\alpha$, the relative enhancement of the cross section is greater and happens at lower values of $v$.

\subsection{Capture Rate}
\label{sec:capture-rate}

An increased capture cross section like in Figure~\ref{fig:capture_cross_section} leads to a higher binary formation rate, and thus to an enhanced merger rate $\mathcal R$. In general, the latter can be computed as
\beq
\mathcal R=\int \dd V\dd M\dd M_*\:n_{\scalebox{0.65}{$M$}}n_{\scalebox{0.65}{$M_*$}}\braket{\sigma\ped{tot}v}\,,
\eeq
where $n_{\scalebox{0.65}{$M$}}$ and $n_{\scalebox{0.65}{$M_*$}}$ are the comoving average number densities of the primary and companion object, respectively, while the integral over $\dd V$ is performed over the volume one is interested in, e.g.\ the Milky Way or the LISA range. The term $\braket{\sigma\ped{tot}v}$ is the capture cross section weighted by some velocity distribution $P(v)$, that is,
\beq
\braket{\sigma\ped{tot}v}=\int \sigma\ped{tot}(v)P(v)v\dd v\,.
\eeq
Depending on the specific astrophysical environments under consideration (e.g.\ globular clusters or active galactic nuclei), a suitable velocity distribution must be chosen, from which the merger rate can be calculated. In practice, however, this approach hides many subtleties such as mass segregation \cite{OLeary:2008myb,Amaro-Seoane:2010dzj}, and the values for the merger rates are very uncertain \cite{Amaro-Seoane:2020zbo}.

\vskip 4pt
Giving a detailed account of these issues is beyond the scope of this work. We can, however, provide an estimate for the increase in the merger rate due to the presence of the cloud, based on the fact that $\mathcal R$ is directly proportional to $\braket{\sigma\ped{tot}v}$. The maximum increase happens when $P(v)$ has most of its support in correspondence of the peak of $\sigma\ped{tot}/\sigma_\slab{gw}$: in that case, one can expect the merger rate to be enhanced by a factor of $\mathcal O(10)-\mathcal O(100)$, depending on the parameters. Any other velocity distribution will give an increase by a factor from 1 up to that maximum value. For the parameters chosen in Figure~\ref{fig:capture_cross_section}, the peak is indeed located at values of $v$ close to the typical velocities found in the center of Milky Way-like galaxies: we can thus expect the rate of events with $q\sim10^{-3}$ to be significantly enhanced. On the other hand, from (\ref{eqn:scaling-cross-section}), we note that the peak shifts to lower values of $v$ when the mass ratio is reduced, hinting to a less significant increase for the rate of EMRIs with $q\ll10^{-3}$.

\section{Ionization and Dynamical Friction}
\label{sec:ionization-dynamical-friction}

When the orbital separation in a binary system is roughly comparable with the size of the superradiant cloud, a strong cloud-binary interaction occurs. This results in a partial destruction of the cloud, with consequent energy loss at the binary's expense. The effect has been studied in \cite{Baumann:2021fkf,Baumann:2022pkl}, where it has been dubbed ``ionization'', for its analogy with the homonymous process in atomic physics.

\vskip 4pt
In Section~\ref{sec:ionization-review}, we briefly review the derivation and the main features of ionization, whose treatment will be extended in Sections~\ref{sec:eccentricity} and \ref{sec:inclination} to more general cases. Then, in Section~\ref{sec:dynamical-friction}, we discuss extensively the interpretation of the backreaction of ionization on the orbit, with particular focus on its relation with the well-known effect of dynamical friction.

\subsection{Ionizing the Cloud}
\label{sec:ionization-review}

Ionization is the partial transfer of the cloud from its starting bound state $\ket{n_b\ell_bm_b}$ to any unbound states $\ket{k;\ell m}$. This process can be mediated by the time-varying gravitational perturbation $V_*(t, \vec r)$ in a binary system. As in Section~\ref{sec:E_lost}, it is useful to decompose the wavefunction as
\beq
\ket{\psi(t)}=c_{n_b\ell_bm_b}(t)\ket{n_b\ell_bm_b}+\int\frac{\dd k}{2\pi}\sum_{\ell, m}c_{k;\ell m}(t)\ket{k;\ell m}\,.
\eeq
Similar to (\ref{eqn:cnlm-b}), the coefficients $c_{k;\ell m}$ can be computed perturbatively as
\beq
c_{k;\ell m}(t)=-i\int_0^{t}\dd t'\braket{k;\ell m|V_*(t',\vec r)|n_b\ell_bm_b}e^{i(\epsilon(k)-\epsilon_b)t'}=i\,\eta\,\frac{1-e^{i(\epsilon(k)-\epsilon_b-g\Omega)t}}{i(\epsilon(k)-\epsilon_b-g\Omega)}\,,
\label{eqn:cnlm-u-ion}
\eeq
where $\eta$ is defined in (\ref{eqn:def-eta}). The last equality only holds on equatorial quasi-circular orbits. In order to obtain it, we exploited the selection rules of the angular integral $I_\Omega$, which hides inside the matrix element $\eta$: of all terms in the perturbation, only those that oscillate with frequency $g\Omega$ survive, where $g=m-m_b$ for co-rotating orbits and $g=m_b-m$ for counter-rotating orbits. When a long-time average of $\abs{c_{k;\ell m}}^2$ is taken, the time-dependent numerator of (\ref{eqn:cnlm-u-ion}) combines with the denominator to produce a delta function:
\beq
\abs{c_{k;\ell m}}^2=2\pi t\,\abs{\eta}^2\,\delta\bigl(\epsilon(k)-\epsilon_b-g\Omega\bigr)\,.
\label{eqn:fermis-golden-rule}
\eeq
Equation (\ref{eqn:fermis-golden-rule}) is nothing more than Fermi's Golden Rule. Summing over all unbound states yields the total ionization rate,
\beq
\frac{\dot M\ped{c}}{M\ped{c}}=-\sum_{\ell,g}\,\frac{\mu\abs{\eta^\floq{g}}^2}{k_\floq{g}}\Theta\bigl(k_\floq{g}^2\bigr)\,,
\label{eqn:rate}
\eeq
where we defined $k_\floq{g}=\sqrt{2\mu(\epsilon_b+g\Omega)}$, as well as the matrix element $\eta^\floq{g}$ of $V_*$ between the states $\ket{k_\floq{g};\ell,m_b\pm g}$ and $\ket{n_b\ell_bm_b}$. Similarly, one can define the rates of energy (``ionization power'') and angular momentum (``ionization torque'') transferred into the continuum as
\begin{align}
\label{eqn:pion}
P\ped{ion}&=\frac{M\ped{c}}\mu\sum_{\ell,g}\,g\Omega\,\frac{\mu\abs{\eta^\floq{g}}^2}{k_\floq{g}}\Theta\bigl(k_\floq{g}^2\bigr)\,,\\
\label{eqn:tion}
\tau\ped{ion}&=\frac{M\ped{c}}\mu\sum_{\ell,g}\,g\,\frac{\mu\abs{\eta^\floq{g}}^2}{k_\floq{g}}\Theta\bigl(k_\floq{g}^2\bigr)\,.
\end{align}

\begin{figure}
\centering
\includegraphics{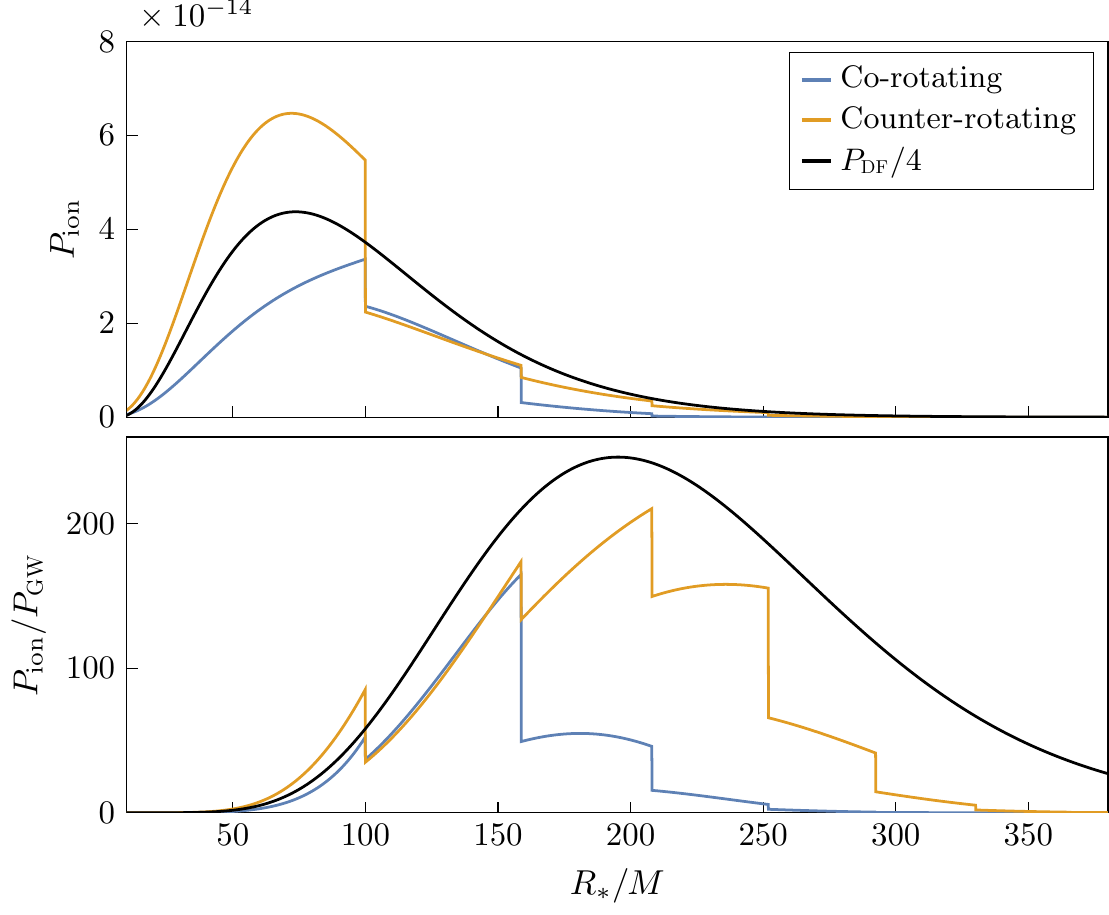}
\caption{Ionization power (\ref{eqn:pion}) as function of the orbital separation $R_*$, for $\alpha=0.2$, $q=10^{-3}$, $M\ped{c}=0.01M$ and a cloud in the $\ket{211}$ state. The top panel shows $P\ped{ion}$ in units where $M=1$. The bottom panel shows the ratio $P\ped{ion}/P_\slab{gw}$. Shown is also $P_\slab{df}$, defined in (\ref{eqn:df}).}
\label{fig:ionization_211}
\end{figure}

\vskip 4pt
When equations (\ref{eqn:rate}), (\ref{eqn:pion}) and (\ref{eqn:tion}) are evaluated numerically, two noteworthy features are found. First, we note that $P\ped{ion}$ is much larger than $P_\slab{gw}$ for a wide range of orbital separations (see bottom panel of Figure~\ref{fig:ionization_211}). This means that the backreaction of ionization dominates over the radiation reaction due to the emission of gravitational waves, which is the main driving force of the inspiral in vacuum. This result holds in a large region of parameter space. In particular, it can be shown that, in the small-$q$ limit,
\beq
\frac{P\ped{ion}}{P_\slab{gw}}=\frac{M\ped{c}}M\alpha^{-5}\mathcal F(\alpha^2R_*/M)\,,
\label{eqn:pion-pgw-scaling}
\eeq
where $\mathcal F$ is a function that only depends on the bound state. The scaling relation (\ref{eqn:pion-pgw-scaling}) allows to quickly adapt the results of Figure~\ref{fig:ionization_211} to any values of choice for the parameters.

\vskip 4pt
The second main feature of ionization are the sharp discontinuities exhibited by $P\ped{ion}$ (as well as by the rate and torque) at separations corresponding to the orbital frequencies
\beq
\Omega^\floq{g}=\frac{\alpha^3}{2gMn_b^2}\,,\qquad g=1,2,3,\ldots
\label{eqn:omega-g}
\eeq
These can be interpreted as \emph{threshold frequencies}, in analogy to the ones found in the photoelectric effect. In our case, because the perturbation is not monochromatic, each different Fourier component produces a different jump. It is important to realize that, while $P\ped{ion}$ is indeed discontinuous in the limit where $\Omega$ is kept fixed, in reality the orbital frequency ramps up as the binary inspirals. As a consequence, the discontinuities are replaced by smooth, although steep, transient oscillating phenomena, thoroughly described in \cite{Baumann:2021fkf}.

\subsection{Interpretation as Dynamical Friction}
\label{sec:dynamical-friction}

As detailed in the Section~\ref{sec:ionization-review}, ionization pumps energy into the scalar field. This must happen at the expense of the binary's total energy, meaning that ionization backreacts on the orbit by inducing an energy loss, or a ``drag force''. The effect peaks roughly when the orbital separation equals the distance at which the cloud is densest, as is clear from the top panel of Figure~\ref{fig:ionization_211}. This conclusion is hardly a surprise. The existence of a drag force acting on an object (in our case, the secondary body of mass $M_*$) that moves through a medium (the cloud) with which it interacts gravitationally is well-established, and goes under the name of \emph{dynamical friction}. In this section, our goal is to give a detailed comparison between ionization and well-known results about dynamical friction, eventually showing that the two effects should be interpreted as one.

\vskip 4pt
Dynamical friction was first studied by Chandrasekhar in \cite{Chandrasekhar:1943ys} for a medium composed of collisionless particles. More recently, results have been found for the motion in an ultralight scalar field \cite{Hui:2016ltb,Traykova:2021dua,Vicente:2022ivh,Traykova:2023qyv,Buehler:2022tmr}, which is relevant for our case. For non-relativistic velocities, the dynamical friction power is found to be
\beq
P_\slab{df}=\frac{4\pi M_*^2\rho}v\bigl(\log(v\mu b\ped{max})+\gamma_\slab{e}\bigr)\,.
\label{eqn:df}
\eeq
We now define the parameters entering (\ref{eqn:df}), as well as highlight all the assumptions behind it.

\begin{enumerate}
\item At large distance from the object of mass $M_*$, the medium is assumed to be uniform with density $\rho$. The velocity $v$ of the object is measured with respect to the asymptotically uniform regions of the medium.
\item The motion of the object is assumed to be uniform and straight. In particular, this implies that its interaction with the medium started an infinitely long time in the past.
\item If the two previous assumptions are taken strictly, the result for $P_\slab{df}$ is logarithmically divergent. The reason is that, in the stationary configuration, the medium forms an infinitely extended wake of overdensity behind the moving body, whose gravitational pull on the object diverges. A regulator is thus introduced: the parameter $b\ped{max}$ sets an upper bound to the impact parameter of the elements of the medium whose interaction with the object is taken into account. The last factor of (\ref{eqn:df}) depends on $b\ped{max}$ (logarithmically), as well as on the mass of the scalar field $\mu$ and the Euler-Mascheroni constant $\gamma_\slab{e}\approx0.577$.
\end{enumerate}

\vskip 4pt
Before applying formula (\ref{eqn:df}) to the case of a gravitational atom in a binary, one must realize that these three points all fail or need modifications: (1) the medium is not uniform and has a finite size; as a consequence, the relative velocity $v$ must be redefined; (2) the object moves in a circle rather than in a straight line; (3) the finiteness of the medium acts as a natural regulator for the divergence of $P_\slab{df}$; as a consequence, the parameter $b\ped{max}$ (which would not be needed in a self-consistent calculation) must be fixed with a suitable choice. Nevertheless, formula (\ref{eqn:df}), as well as similar ones for other kinds of media, are routinely applied in similar astrophysical contexts \cite{Eda:2014kra,Macedo:2013qea,Barausse:2014tra,Zhang:2019eid,Kavanagh:2020cfn,Kim:2022mdj}, with the expectation that they capture the correct dependence on the parameters and provide a result which is correct up to factors of $\mathcal O(1)$.

\vskip 4pt
Let us now evaluate (\ref{eqn:df}) in our case, adopting choices for the various parameters that are common in the literature. We set $\rho$ equal to the local density of the cloud at the companion's position, $\rho=M\ped{c}\abs{\psi(\vec R_*)}^2$; we fix $v$ equal to the orbital velocity, $v=\sqrt{(1+q)M/R_*}$; finally, we choose $b\ped{max}=R_*$. Note that these choices are, strictly speaking, mutually inconsistent: for example, we are considering impact parameters as large as the size of the orbit, but ignoring that over such distance the cloud's density varies significantly compared to its local value.

\begin{figure}
\centering
\includegraphics{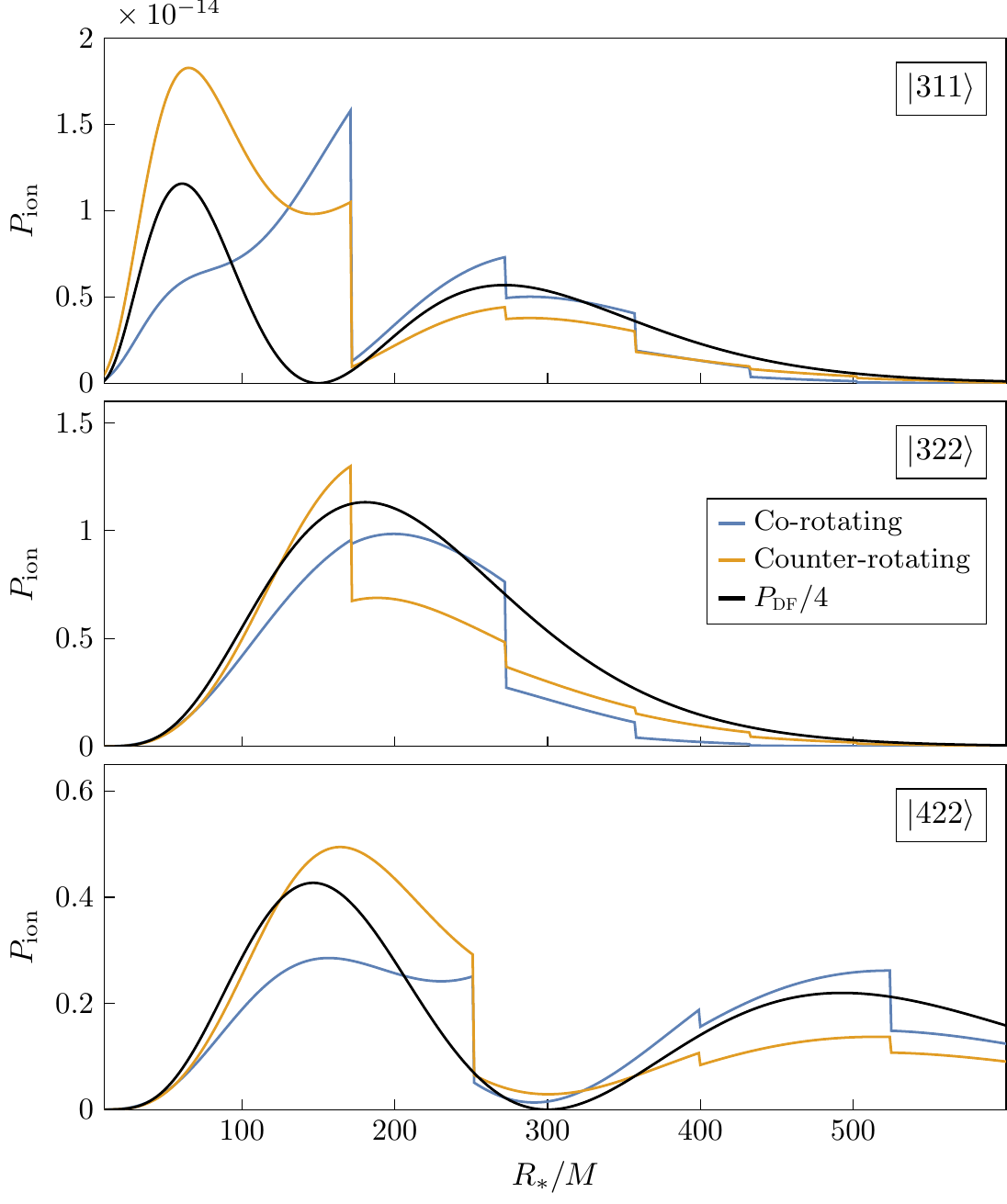}
\caption{Comparison of $P_\slab{df}$ (divided by 4 for clarity) with $P\ped{ion}$, for clouds in the states $\ket{311}$, $\ket{322}$ and $\ket{422}$. All the parameters and the units are the same as in Figure~\ref{fig:ionization_211}.}
\label{fig:dynamical-friction}
\end{figure}

\vskip 4pt
With these assumptions, we calculate $P_\slab{df}$ and compare it to $P\ped{ion}$ in Figures~\ref{fig:ionization_211} and \ref{fig:dynamical-friction}, for a selection of states $\ket{n_b\ell_bm_b}$. In all cases, $P_\slab{df}$ turns out to be a factor of a few larger than $P\ped{ion}$; for the sake of a better visual comparison, we plotted the fourth part of $P_\slab{df}$ instead. The various states have been selected not necessarily because they are expected to be populated by superradiance, but simply to exhibit the comparison between $P_\slab{df}$ and $P\ped{ion}$ on clouds with different profiles. Clearly, $P_\slab{df}$ possesses no discontinuities and, with the assumed values of its parameters, its value does not depend on the orientation of the orbit. In all cases, nevertheless, the two quantities have roughly the same overall shape, generally peaking in correspondence with the densest regions of the cloud and having minima elsewhere. This conclusion does not depend on the chosen values of the parameters: by plugging the assumed values of $\rho$, $v$ and $b\ped{max}$ in (\ref{eqn:df}), it is possible to show that the ratio $P_\slab{df}/P_\slab{gw}$ has exactly the same scaling as $P\ped{ion}/P_\slab{gw}$, given in (\ref{eqn:pion-pgw-scaling}). This means that the ratio $P_\slab{df}/P\ped{ion}$ is universal, and roughly equal to a constant of $\mathcal O(1)$.

\vskip 4pt
Having demonstrated that $P_\slab{df}$ and $P\ped{ion}$ always give the same result, modulo the expected corrections of $\mathcal O(1)$ due to the ambiguities in fixing the parameters entering $P_\slab{df}$, we now briefly discuss, on theoretical grounds, in what sense dynamical friction must be interpreted as the backreaction of ionization. One way to derive $P_\slab{df}$ is to first solve the Schrödinger equation for the Coulomb scattering of the scalar field off the moving object, and then perform a surface integral of (some component of) the energy-momentum tensor of the medium \cite{Hui:2016ltb}. By Newton's third law, the drag force on the moving body is equal to the flux of momentum carried by the medium around it. On the other hand, the physical mechanism behind ionization, as well as the derivation of the result, is basically the same. Due to different boundary conditions, bound states carry no energy-momentum flux at infinity, while unbound states do. We solve perturbatively the Schrödinger equation and determine the rate at which the latter are populated: this defines~$P\ped{ion}$.

\vskip 4pt
The main physical difference between the two cases is the initial, unperturbed state of the medium: unbound for $P_\slab{df}$, bound around the larger object for $P\ped{ion}$. The finite energy jump that separates each bound state from the continuum is the cause of the discontinuities observed in $P\ped{ion}$ but not in $P_\slab{df}$. In this sense, we can say that ionization is sensitive to both local properties of the cloud (as it correlates with its density) and global ones (such as the bound states' spectrum) and is nothing but a self-consistent calculation of dynamical friction for the gravitational atom.

\section{Ionization and Eccentricity}
\label{sec:eccentricity}

A binary that forms via dynamical capture, as discussed in Section~\ref{sec:capture}, is initially characterized by very eccentric orbits. Studies of later stages of the inspiral, when the gravitational wave signal is stronger and the impact of the cloud's ionization becomes more relevant, have instead focused on quasi-circular orbits \cite{Takahashi:2021yhy,Baumann:2021fkf,Baumann:2022pkl,Cole:2022fir}. Most work on resonant transitions also made this same simplifying assumption \cite{Baumann:2018vus,Zhang:2018kib,Zhang:2019eid,Baumann:2019ztm,Takahashi:2021eso,Tong:2022bbl,Takahashi:2023flk}, with only \cite{Berti:2019wnn} considering non-zero eccentricity, at a time where, however, some physical aspects of the problem were not yet completely understood.

\vskip 4pt
In this section, we relax the assumption of circular orbits, generalizing the treatment of ionization to arbitrary eccentricity. We then discuss the evolution of eccentricity due to ionization and emission of GWs, explaining under what conditions the assumption of quasi-circular orbits is justified. However, we still assume for simplicity that the binary lies in the equatorial plane of the cloud: this assumption will be relaxed in Section~\ref{sec:inclination}.

\subsection{Ionization Power and Torque}

As reviewed in Section~\ref{sec:ionization-review}, neglecting the short transient phenomena that happen around the frequencies given in (\ref{eqn:omega-g}), the ionization rates can be found by applying Fermi's Golden Rule to a non-evolving orbit, which requires computing the matrix element 
\beq
\braket{k;\ell m|V_*(t,\vec r)|n_b\ell_bm_b}=-\sum_{\ell_*,m_*}\frac{4\pi\alpha q}{2\ell_*+1}Y_{\ell_*m_*}\biggl(\frac\pi2,\varphi_*\biggr)I_rI_\Omega\,.
\label{eqn:matrix-element-ionization}
\eeq
In the case of a circular orbit, the calculation is simplified by the fact that not only $I_\Omega$, but also $I_r$ is constant in time. The only time dependence of (\ref{eqn:matrix-element-ionization}) is then encoded in the spherical harmonics, each of which oscillates with a definite frequency, because $\varphi_*=\Omega t$ on circular orbits. This allows one to extract analytically the expression of the Fourier coefficient of the matrix element corresponding to a given oscillation frequency $g\Omega$.

\vskip 4pt
On an eccentric Keplerian orbit, the separation $R_*$ and the angular velocity $\dot\varphi_*$ vary with time. A useful parametrization is given in terms of the \emph{eccentric anomaly} $E$:
\beq
R_*=a(1-\varepsilon\cos E)\,,\qquad(1-\varepsilon)\tan^2\frac{\varphi_*}2=(1+\varepsilon)\tan^2\frac{E}2\,,
\eeq
where $a$ is the semi-major axis and $\varepsilon$ is the eccentricity. The eccentric anomaly as function of time must then be found by solving numerically Kepler's equation,
\beq
\Omega t=E-\varepsilon\sin E\,.
\eeq
The matrix element is thus an oscillating function with period $\Omega$, which we can expand in a Fourier series as in (\ref{eqn:def-eta}),
\beq
\braket{k;\ell m|V_*(t,\vec r)|n_b\ell_bm_b}=\sum_{f\in\mathbb{Z}} \eta^\floq{f}e^{-if\Omega t}\,.
\label{eqn:fourier}
\eeq
If $k=\sqrt{2\mu(\epsilon_b+g\Omega)}\equiv k_\floq{g}$, Fermi's Golden Rule tells us that the only term of (\ref{eqn:fourier}) that gives a non-zero contribution to the ionization rate is the one that oscillates with a frequency equal to the energy difference between the two states, that is, the one with $f=g$. By comparison with equation (3.28) of \cite{Baumann:2021fkf}, the ionization rate is
\beq
\frac{\dot M\ped{c}}{M\ped{c}}=-\sum_{\ell, m,g}\,\frac{\mu\abs{\eta^\floq{g}}^2}{k_\floq{g}}\Theta\bigl(k_\floq{g}^2\bigr)\,,
\label{eqn:rate-eccentric}
\eeq
where the sum runs over all continuum states of the form $\ket{k_\floq{g};\ell m}$. The Fourier coefficients $\eta^\floq{g}$ have an implicit dependence on $k_\floq{g}$ as well as on the orbital parameters. Similarly, the ionization power and torque (along the central BH's spin) are\footnote{Equations (\ref{eqn:rate-eccentric}), (\ref{eqn:pion-eccentric}) and (\ref{eqn:tion-eccentric}) are very similar to (\ref{eqn:rate}), (\ref{eqn:pion}) and (\ref{eqn:tion}). The difference is that now $\eta^\floq{g}$ no longer vanishes when $g\ne\pm(m-m_b)$, so we need to sum over $m$ and $g$ independently.}
\begin{align}
\label{eqn:pion-eccentric}
P\ped{ion}&=\frac{M\ped{c}}\mu\sum_{\ell, m,g}\,g\Omega\,\frac{\mu\abs{\eta^\floq{g}}^2}{k_\floq{g}}\Theta\bigl(k_\floq{g}^2\bigr)\,,\\
\label{eqn:tion-eccentric}
\tau\ped{ion}&=\frac{M\ped{c}}\mu\sum_{\ell, m,g}\,(m-m_b)\,\frac{\mu\abs{\eta^\floq{g}}^2}{k_\floq{g}}\Theta\bigl(k_\floq{g}^2\bigr)\,.
\end{align}

\vskip 4pt
An important difference with respect to the circular case is that it is no longer true that $P\ped{ion}=\Omega\tau\ped{ion}$. The equality held because, in that case, $Y_{\ell_*m_*}(\pi/2,\Omega t)$ was the only time-dependent term of (\ref{eqn:matrix-element-ionization}). This spherical harmonic oscillates with frequency $m_*\Omega$, which is fixed by the angular selection rules in $I_\Omega$ to be $\pm(m-m_b)\Omega$, depending on the orbit's orientation. For $\varepsilon>0$, instead, the factors entering $P\ped{ion}$ and $\tau\ped{ion}$ are independent. As we will see in Section~\ref{sec:evolution-e}, the evolution of the eccentricity will be determined by the ratio $\tau\ped{ion}/P\ped{ion}$.

\subsection{Numerical Evaluation}
\label{sec:numerics-e}

The complexity of expressions (\ref{eqn:rate-eccentric}), (\ref{eqn:pion-eccentric}) and (\ref{eqn:tion-eccentric}) is hidden in the Fourier coefficients $\eta^\floq{g}$, which we evaluate numerically. Their expression,
\beq
\eta^\floq{g}=\int_0^{2\pi/\Omega}\braket{k;\ell m|V_*(t,\vec r)|n_b\ell_bm_b}e^{ig\Omega t}\dd t\supset\int_0^{\pi/\Omega}\cos(m_*\phi_*+g\Omega t)I_r(t)\dd t\,,
\eeq
contains the overlap integral $I_r$ nested inside a time integral, as we made manifest in the last term, where we neglected all time-independent coefficients. In order to improve the convergence of the numerical routine, we write the time integrals as
\beq
\int_0^{\pi/\Omega}\biggl[\cos[m_*(\phi_*-\Omega t)]I_r(t)\cos[(m_*+g)\Omega t]-\sin[m_*(\phi_*-\Omega t)]I_r(t)\sin[(m_*+g)\Omega t]\biggr]\dd t\,.
\eeq
The monochromatic oscillatory term $\cos[(m_*+g)\Omega t]$ multiplies a function, $\cos[m_*(\phi_*-\Omega t)]I_r(t)$, whose $\varepsilon\to0$ limit is time-independent (and similarly for the second term, replacing the cosine with a sine). This form makes it therefore particularly convenient to perform the integration using a routine optimized for definite Fourier integrals, as the $\varepsilon\to0$ limit is expected to be numerically smooth and should recover the result for circular orbits. The task is nevertheless computationally expensive: increasing the eccentricity requires to extend the sum to a larger number of final states to achieve a good numerical precision; moreover, the convergence of the integrals starts to degrade for $\varepsilon\gtrsim0.7$.

\begin{figure}
\centering
\includegraphics{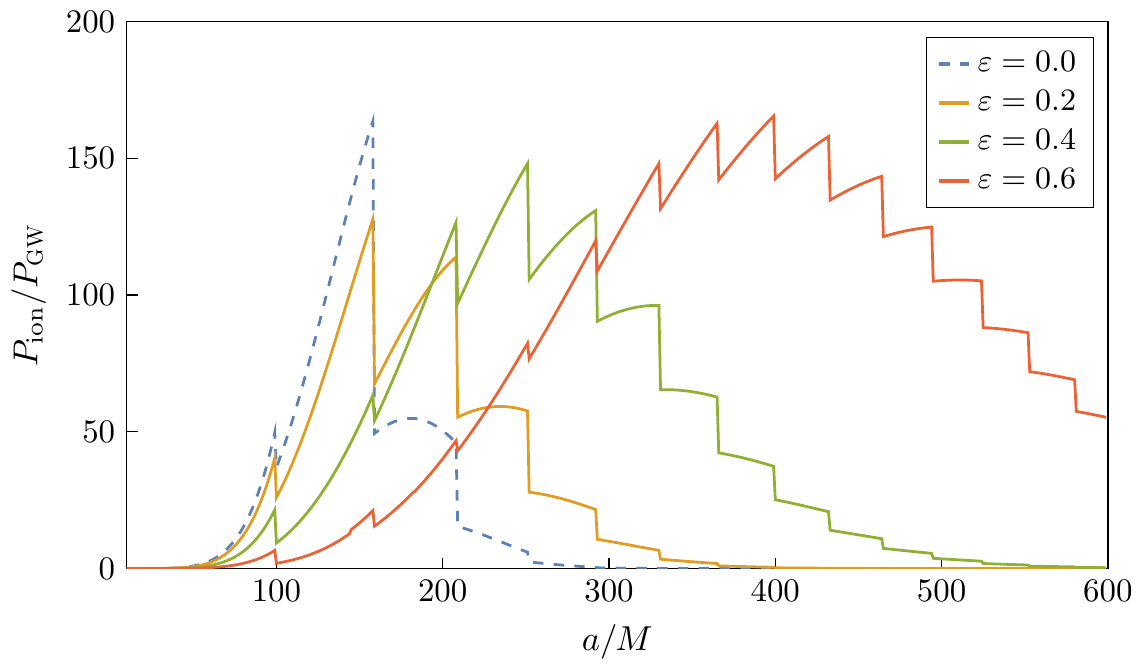}
\caption{Ionization power (\ref{eqn:pion-eccentric}) for different values of the eccentricity $\varepsilon$, as function of the semi-major axis $a$. The values are normalized by $P_\slab{gw}$, the average power emitted in gravitational waves on a correspondingly eccentric orbit, and are computed for $\alpha=0.2$, $q=10^{-3}$, $M\ped{c}=0.01M$, a cloud in the $\ket{211}$ state, and co-rotating equatorial orbits.}
\label{fig:P_ion_P_GW_eccentricity_211}
\end{figure}

\vskip 4pt
In Figure~\ref{fig:P_ion_P_GW_eccentricity_211}, we show $P\ped{ion}$ as function of the semi-major axis $a$, for different values of the eccentricity $\varepsilon$. We normalize the result by $P_\slab{gw}$, which itself depends on the eccentricity and is defined as an orbit-averaged value. The characteristic discontinuities of $P\ped{ion}$ remain at the same positions, as they are determined by the value of the orbital frequency (\ref{eqn:omega-g}), which is only a function of $a$. On the other hand, the peak of the curve shifts to larger values of $a$ for increasing $\varepsilon$. This implies that the effect of ionization is felt earlier on eccentric binaries. Similar calculations and considerations hold for the ionization rate (\ref{eqn:rate-eccentric}) and the torque (\ref{eqn:tion-eccentric}). While in Figure~\ref{fig:P_ion_P_GW_eccentricity_211} we assumed the cloud to be in the $\ket{211}$ state, we show the same results for $\ket{322}$ in Appendix~\ref{sec:ion322}.

\subsection{Evolution of Eccentricity}
\label{sec:evolution-e}

We now have all the ingredients to compute the backreaction of ionization on eccentric orbits. While a detailed solution of the evolution of the system should include the accretion of matter on the companion (if it is a BH) and the mass loss of the cloud \cite{Baumann:2021fkf}, as well as its self gravity \cite{Ferreira:2017pth,Hannuksela:2018izj}, to first approximation we may neglect all of these effects. With respect to the case of circular orbits, the evolution of the semi-major axis does not present new insightful features: we can determine it with the energy balance equation alone,
\beq
\frac\dd{\dd t}\biggl(-\frac{qM^2}{2a}\biggr)=-P\ped{ion}-P_\slab{gw}\,,
\label{eqn:energy-balance-eccentric}
\eeq
where $P_\slab{gw}$ is defined in (\ref{eq:p_gw}) and $P\ped{ion}$ has the effect of making $a$ decrease faster than expected in vacuum. Much less trivial is the evolution of the eccentricity. In order to find it, we need the balance of angular momentum,
\beq
\frac\dd{\dd t}\sqrt{\frac{q^2M^3}{1+q}a(1-\varepsilon^2)}=-\tau\ped{ion}-\tau_\slab{gw}\,,
\label{eqn:L-balance-eccentric}
\eeq
where $\tau_\slab{gw}$ is defined in (\ref{eq:tau_gw}). This equation can then be used together with (\ref{eqn:energy-balance-eccentric}) to find $\dd\varepsilon/\dd t$.

\begin{figure}
\centering
\includegraphics{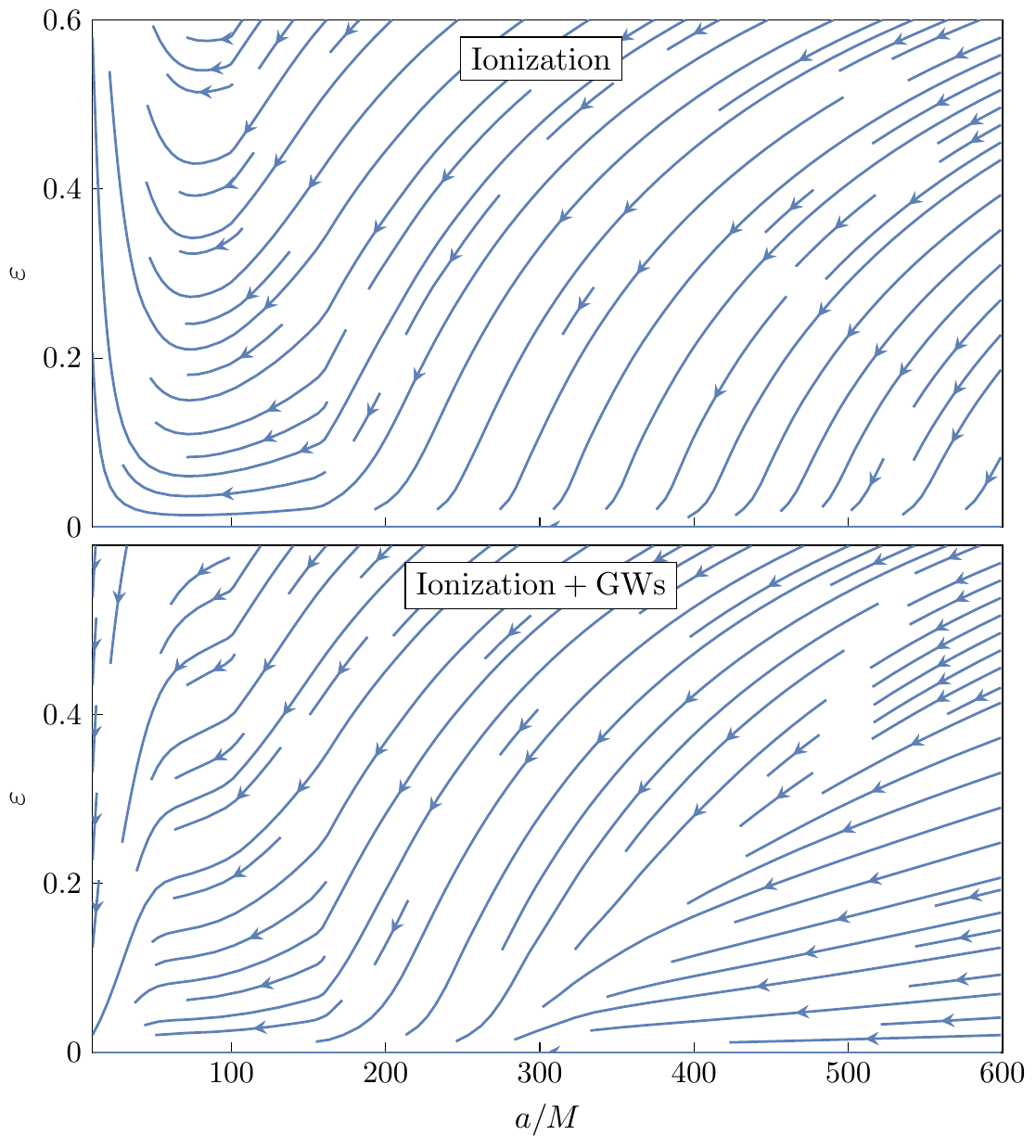}
\caption{Numerical solutions to (\ref{eqn:de/da}), for various different initial values of the semi-major axis and the eccentricity. The top panel neglects $P_\slab{gw}$ and $\tau_\slab{gw}$, while the bottom panel shows the solution to the complete equation. The values of the parameters and the orientation of the orbit are the same as in Figure~\ref{fig:P_ion_P_GW_eccentricity_211}.}
\label{fig:streamplot_eccentricity_unified}
\end{figure}

\vskip 4pt
The most pressing question is perhaps whether ionization acts to reduce or increase the binary's eccentricity. Besides being an interesting question per se, it is necessary to justify (or disprove) the assumption of quasi-circular orbits adopted in a number of previous works, such as \cite{Baumann:2021fkf,Cole:2022fir}. It is useful to combine (\ref{eqn:energy-balance-eccentric}) and (\ref{eqn:L-balance-eccentric}) as
\beq
\frac{\dd\varepsilon}{\dd(a/M)}=\frac{1-\varepsilon^2}{2\varepsilon(a/M)}-\frac{\sqrt{(1-\varepsilon^2)(1+q)}}{2\varepsilon(a/M)^{5/2}M}\frac{\tau\ped{ion}+\tau_\slab{gw}}{P\ped{ion}+P_\slab{gw}}\,,
\label{eqn:de/da}
\eeq
which allows to numerically integrate the eccentricity $\varepsilon$ as function of the semi-major axis $a$. We do this in Figure~\ref{fig:streamplot_eccentricity_unified}, where several curves corresponding to different initial values of $a$ and $\varepsilon$ are shown. In the top panel, we neglect $P_\slab{gw}$ and $\tau_\slab{gw}$ in (\ref{eqn:de/da}), while in the bottom panel we solve the full equation. Generally speaking, the binary undergoes circularization under the combined effect of ionization and gravitational wave emission. Nevertheless, when gravitational waves are neglected, for small enough $a$ the binary can experience eccentrification.

\vskip 4pt
This interesting behaviour has an insightful qualitative explanation. The density profile of the $\ket{211}$ state, shown in Figure~\ref{fig:E_lost_211}, has a maximum at a certain radius and goes to zero at the center and at infinity. Suppose that the companion is on a very eccentric orbit with semi-major axis larger than the size of the cloud, so that the density of the cloud at periapsis is much higher than at apoapsis. According to the interpretation as dynamical friction laid down in Section~\ref{sec:dynamical-friction}, the drag force experienced at periapsis will thus be much stronger than the one at apoapsis. To approximately model the fact that most of the energy loss is concentrated at the periapsis, we may imagine that the orbiting body receives a ``kick'' every time it passes through the periapsis, with the rest of the orbit being unperturbed. This way, the periapsis of successive orbits stays unchanged, while the apoapsis progressively reduces orbit by orbit: in other words, the binary is circularizing. Conversely, suppose that the semi-major axis is smaller than the size of the cloud. The situation is now reversed: the periapsis will be in a region with lower density, and successive kicks at the apoapsis will eccentrify the binary.

\vskip 4pt
The transition between circularization and eccentrification in the top panel of Figure~\ref{fig:streamplot_eccentricity_unified} happens indeed at a distance comparable with the size of the cloud, supporting the qualitative interpretation of the phenomenon. As is well-known, the emission of gravitational waves has a circularizing effect on binary systems. Indeed, when they are taken into account, the eccentrifying effect of ionization at small values of $a$ is reduced, especially for $a\to0$, where $P_\slab{gw}\gg P\ped{ion}$. It is worth noting, however, that while in Figure~\ref{fig:streamplot_eccentricity_unified} only circularization is allowed after the addition of GWs, it is in principle possible that part of the eccentrifiying effect survives, depending on the parameters (for example, a high enough mass of the cloud would guarantee a ``region'' of~eccentrification). An example of this is given in Appendix \ref{sec:appA}.

\section{Ionization and Inclination}
\label{sec:inclination}

Gravitational atoms are not spherically symmetric systems. Not only must the central BH be spinning around its axis to trigger superradiance, but the cloud itself is necessarily generated in a state with non-zero angular momentum, implying that it must have a non-trivial angular structure. Its impact on the evolution of a binary system will therefore depend on the inclination $\beta$ of the orbital plane with respect to equatorial plane defined by the spins of central BH and its cloud.

\vskip 4pt
To the best of our knowledge, no study of gravitational atoms in binaries has so far considered non-equatorial orbits. In this section, we will relax this assumption for the first time, by extending the treatment of ionization to the full range $0\leq\beta\leq\pi$. Precession of the orbital plane and evolution of the inclination angle will then be discussed. Motivated by the results of Section~\ref{sec:eccentricity}, in this section we will assume for simplicity that the orbits are quasi-circular.

\begin{figure}
\centering
\includegraphics[scale=1.2]{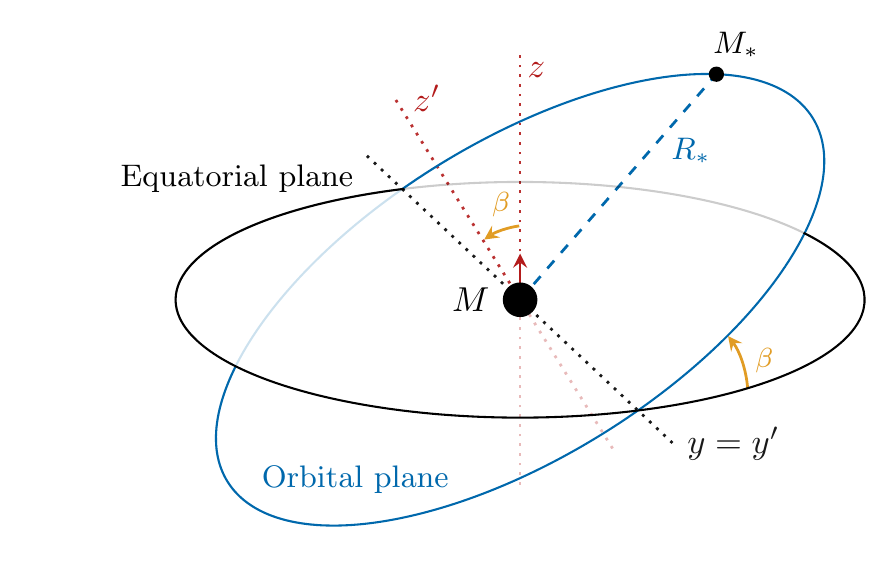}
\caption{Diagram of the coordinates used to describe inclined orbits. The orbital plane is obtained by rotating the equatorial plane by an angle $\beta$ around the $y$ axis.}
\label{fig:InclinedOrbit}
\end{figure}

\vskip 4pt
Before detailing the calculation, it is useful to state our conventions clearly. With reference to Figure~\ref{fig:InclinedOrbit}, we align the $z$ axis with the BH's spin and the $y$ axis with the intersection of the equatorial plane with the orbital plane. We use the $z$-$y$-$z$ convention for the Euler angles, so that the Euler angle $\beta$ is defined in the $x$-$z$ plane and is identified with the orbital inclination. The axes $x'$, $y'$ and $z'$, instead, will be aligned with the binary's orbit, with $y'\equiv y$.

\subsection{Ionization Power and Torque}

The most obvious way to compute the ionization power and torque on an inclined orbit is to simply evaluate the perturbation (\ref{eqn:V_star}) accordingly. As we assume a constant $R_*$, the only term that depends on the inclination angle $\beta$ is the spherical harmonic $Y_{\ell_*m_*}(\theta_*(t),\varphi_*(t))$. This can be written as \cite{wigner}
\beq
Y_{\ell_*m_*}(\theta_*(t),\varphi_*(t))=\sum_{g=-\ell_*}^{\ell_*}d_{m_*,-g}^\floq{\ell_*}(\beta)Y_{\ell_*,-g}\biggl(\frac\pi2,0\biggr)e^{-ig\Omega t}\,,
\label{eqn:rotated-Y}
\eeq
where $d_{m'm}^\floq{j}(\beta)$ is a Wigner small $d$-matrix, that in our conventions reads
\beq
d_{m'm}^\floq{j}(\beta)=\mathcal N\sum_{s=s\ped{min}}^{s\ped{max}}\frac{(-1)^{m'-m+s}\left(\cos\frac\beta2\right)^{2j+m-m'-2s}\left(\sin\frac\beta2\right)^{m'-m+2s}}{(j+m-s)!s!(m'-m+s)!(j-m'-s)!}\,,
\eeq
with $s\ped{min}=\max(0,m-m')$, $s\ped{max}=\min(j+m,j-m')$ and the normalization factor given by $\mathcal N=\sqrt{(j+m')!(j-m')!(j+m)!(j-m)!}$. As the expansion (\ref{eqn:rotated-Y}) separates the various monochromatic components, it is possible to proceed in a similar fashion to Fermi's Golden Rule, i.e.\ by only keeping the terms that survive a long-time average in first-order perturbation theory. In this way, we can find the total energy and angular momentum in the continuum. In order to find the ionization power and torque, however, one must subtract the energy and angular momentum remaining in the bound state, and this approach hides an important subtlety, as we will discuss now.

\vskip 4pt
On equatorial orbits, only one of the $2\ell_*+1$ terms in (\ref{eqn:rotated-Y}) is not zero and the binary's gravitational perturbation generates a transfer from the bound state $\ket{n_b\ell_bm_b}$ to continuum states. When a rotation is applied to the orbit, however, the new terms appearing in (\ref{eqn:rotated-Y}) can mediate transitions to the entire set of quasi-degenerate states $\ket{n_b\ell_bm'}$, with $m'\ne m_b$ (although not to states $\ket{n_b\ell'm'}$ with $\ell'\ne\ell_b$, as rotations do not mix different values of $\ell$). In other words, the quasi-degenerate states $\ket{n_b\ell_bm'}$ can be excited. The amount by which this happens is important in determining the ionization torque, as this is determined by the \emph{total} angular momentum carried by the scalar field, be it in continuum or bound states.

\vskip 4pt
In order to consistently describe the phenomenon, it is useful to take another approach and apply a rotation to the bound state, transforming it into a mixture of quasi-degenerate states,
\beq
\ket{n_b\ell_bm_b}\to\sum_{m'=-\ell_b}^{\ell_b}d_{m_bm'}^\floq{\ell_b}(\beta)\ket{n_b\ell_bm'}\,,
\label{eqn:rotated-state}
\eeq
which will then be perturbed by an equatorial orbit. It is important to realize that only in the limit where the Hamiltonian is invariant under rotations this approach is expected to be equivalent to the one where the orbit is rotated instead. Isotropy is only restored in the limit of vanishing BH spin, $\tilde a\to0$, while at finite spin a hyperfine splitting between the states, proportional to $\tilde a$, is present. Assuming that the ionization rate, power and torque for a given inclination angle $\beta$ are continuous in the limit $\tilde a\to0$, the two approaches will become approximately equivalent for sufficiently small BH spin. We can translate this observation into a requirement on the orbital separation by noting that there are only two relevant frequencies in the problem: the orbital frequency $\Omega=\sqrt{M(1+q)/R_*^3}$ and the hyperfine splitting $\Delta\epsilon$, which can be found from (\ref{eq:eigenenergy}). By requiring $\Delta\epsilon\ll\Omega$, we get
\beq
R_*\ll\biggl(\frac{\ell_b(\ell_b+1 / 2)(\ell_b+1)}{2\mu \tilde a\alpha^5\abs{m_b-m'} }\biggr)^{2/3}M^{1/3}(1+q)^{1/3}n_b^2\,.
\label{eqn:hyperfine_R_*}
\eeq
In other words, the rest of the discussion in this section, as well as all the results presented, will only be valid at orbital separations much smaller than the distance of the hyperfine resonance, defined by (\ref{eqn:hyperfine_R_*}). This is a well-justified assumption, as this region of space is parametrically larger than the ``Bohr'' region, where ionization peaks; for typical parameters, it is also larger than the region where $P\ped{ion}/P_\slab{gw}$ has most of its support.

\vskip 4pt
Let us therefore assume that the cloud is in the mixed state given in (\ref{eqn:rotated-state}) and consider its perturbation by an equatorial orbit. Because the matrix elements oscillate monochromatically, at fixed momentum $k$ and angular momentum $\ell$ of the final state, a state $\ket{n_b\ell_bm'}$ can only be ionized towards $\ket{k_\floq{g};\ell,m'+g}$, where $g\Omega=k_\floq{g}^2/(2\mu)-\epsilon_b$. Each of the $2\ell_b+1$ states appearing in (\ref{eqn:rotated-state}) is therefore ionized ``independently'', meaning that no interference terms are generated. We can thus find the total ionization rate, power and $z$ component of the torque by simply adding the contributions from all the $2\ell_b+1$ bound states:
\begin{align}
\label{eqn:rate-inclined}
\frac{\dot M\ped{c}}{M\ped{c}}&=-\sum_{\ell, g,m'}\,\bigl(d_{m_bm'}^\floq{\ell_b}(\beta)\bigr)^2\ \frac{\mu\big|\eta^\floq{g}_{m'}\big|^2}{k_\floq{g}}\Theta\bigl(k_\floq{g}^2\bigr)\,,\\
\label{eqn:pion-inclined}
P\ped{ion}&=\frac{M\ped{c}}\mu\sum_{\ell, g,m'}\,g\Omega\,\bigl(d_{m_bm'}^\floq{\ell_b}(\beta)\bigr)^2\ \frac{\mu\big|\eta^\floq{g}_{m'}\big|^2}{k_\floq{g}}\Theta\bigl(k_\floq{g}^2\bigr)\,,\\
\label{eqn:tauz'-inclined}
\tau\ped{ion}^{z'}&=\frac{M\ped{c}}\mu\sum_{\ell, g,m'}\,g\,\bigl(d_{m_bm'}^\floq{\ell_b}(\beta)\bigr)^2\ \frac{\mu\big|\eta^\floq{g}_{m'}\big|^2}{k_\floq{g}}\Theta\bigl(k_\floq{g}^2\bigr)\,.
\end{align}
In these expressions, we denoted by $\eta^\floq{g}_{m'}$ the matrix element of the perturbation $V_*$ between the states $\ket{k_\floq{g};\ell,m'+g}$ and $\ket{n_b\ell_bm'}$, with the same relation between $k_\floq{g}$ and $g$ as above. Note that it is very easy to go from the expression for $\dot M\ped{c}/M\ped{c}$ to the ones for $P\ped{ion}$ and $\tau\ped{ion}^{z'}$: because the states $\ket{n_b\ell_bm'}$ and $\ket{k;\ell m}$ are simultaneously eigenstates of the energy and of the $z'$ component of the angular momentum, we simply weight each term by the corresponding difference of the eigenvalues: $g\Omega$ for the energy, and $g$ for the angular momentum.

\vskip 4pt
The component $\tau\ped{ion}^{z'}$ of the torque given in (\ref{eqn:tauz'-inclined}) is relative to the axis $z'$, which is orthogonal to the orbital plane. In principle, however, there may also be components that lie in the orbital plane. Our basis does not include eigenstates of the $x'$ or $y'$ components of the angular momentum, meaning that finding the expressions for $\tau\ped{ion}^{x'}$ and $\tau\ped{ion}^{y'}$ requires a little more attention. First, remember that the matrix elements of the angular momentum operator are, in the Condon--Shortley convention, given by
\beq
L_\pm\ket{\ell,m}=\sqrt{\ell(\ell+1)-m(m\pm1)}\ket{\ell,m\pm1}\,,
\eeq
where
\beq
L_{x'}=\frac{L_++L_-}2\,,\qquad L_{y'}=\frac{L_+-L_-}{2i}\,.
\eeq
The time derivative of the angular momentum contained in the continuum states is thus
\beq
\tau\ped{out}^\pm=\frac{M\ped{c}}\mu\frac\dd{\dd t}\int\frac{\dd k}{2\pi}\sum_{\ell,m}\sqrt{\ell(\ell+1)-m(m\pm1)}\,c_{k;\ell,m\pm1}^*c_{k;\ell m}\,.
\eeq
Fermi's Golden Rule only gives the result for $\dd \abs{c_{k;\ell m}}^2/\dd t$. We thus have to go one step back and remember how the amplitudes evolve to first order in perturbation theory:
\beq
c_{k;\ell m}(t)=i\, d_{m_bm'}^\floq{\ell_b}(\beta)\,\eta_{m'}\,\frac{1-e^{i(\epsilon(k)-\epsilon_b-g\Omega)t}}{i(\epsilon(k)-\epsilon_b-g\Omega)}\,,
\label{eqn:c_kellm}
\eeq
where $\eta_{m'}$ is the matrix element of $V_*$ between $\ket{k;\ell m}$ and $\ket{n_b\ell_bm'}$. The time-dependent part of (\ref{eqn:c_kellm}) only depends on $g$, and is thus the same for $c_{k;\ell m}$ and $c_{k;\ell,m\pm1}^*$, while the prefactor differs. We can thus still apply Fermi's Golden Rule, the only difference with the previous cases being that the prefactor $\bigl(d_{m_bm'}^\floq{\ell_b}(\beta)\bigr)^2\big|\eta^\floq{g}_{m'}\big|^2$ will be replaced by its corresponding mixed product. This gives
\begin{align}
\label{eqn:tauoutx'-inclined}
\tau\ped{out}^{x'}&=\frac{M\ped{c}}\mu\sum_{\ell,g,m'}\,\sqrt{\ell(\ell+1)-m(m+1)}\,d_{m_b,m'+1}^\floq{\ell_b}(\beta)d_{m_bm'}^\floq{\ell_b}(\beta)\frac{\mu\,\eta^\floq{g}_{m'+1}\eta^\floq{g}_{m'}}{k_\floq{g}}\Theta\bigl(k_\floq{g}^2\bigr)\,,\\
\label{eqn:tauouty'-inclined}
\tau\ped{out}^{y'}&=0\,.
\end{align}
The vanishing of $\tau\ped{out}^{y'}$ is a consequence of the fact that, in our conventions, both the couplings $\eta_{m'}^\floq{g}$ and the Wigner matrices $d_{m'}^\floq{g}$ are real.

\vskip 4pt
Having computed $\tau\ped{out}^{x'}$ and $\tau\ped{out}^{y'}$, we still need to find the corresponding quantity for the angular momentum contained in the bound states,
\beq
\tau\ped{in}^\pm=\frac{M\ped{c}}\mu\frac\dd{\dd t}\sum_{m'}\sqrt{\ell_b(\ell_b+1)-m'(m'\pm1)}\,c^*_{n_b\ell_b,m'\pm1}c_{n_b\ell_bm'}\,.
\eeq
In this case, the evolution of the amplitude of each state is determined by its own ionization rate, via the requirement of unitarity (again, to first order in perturbation theory):
\beq
c_{n_b\ell_bm'}(t)=d_{m_bm'}^\floq{\ell_b}(\beta)\biggl(1-t\sum_{\ell, g}\,\frac{\mu\big|\eta^\floq{g}_{m'}\big|^2}{2k_\floq{g}}\Theta\bigl(k_\floq{g}^2\bigr)\biggr)\,.
\eeq
We thus find
\beq
\tau\ped{in}^\pm=-\sum_{\ell, g,m'}\sqrt{\ell_b(\ell_b+1)-m'(m'\pm1)}\,d_{m_b,m'\pm1}^\floq{\ell_b}(\beta)d_{m_bm'}^\floq{\ell_b}(\beta)\biggl(\frac{\mu\big|\eta^\floq{g}_{m'\pm1}\big|^2}{2k_\floq{g}}+\frac{\mu\big|\eta^\floq{g}_{m'}\big|^2}{2k_\floq{g}}\biggr)\Theta\bigl(k_\floq{g}^2\bigr)\,,
\eeq
which can be expressed as
\begin{align}
\label{eqn:tauinx'-inclined}
\tau\ped{in}^{x'}&=-\frac{M\ped{c}}\mu\sum_{\ell, g,m'}J^\floq{\ell_b}_{m_bm'}(\beta)\,d_{m_bm'}^\floq{\ell_b}(\beta)\,\frac{\mu\big|\eta^\floq{g}_{m'}\big|^2}{k_\floq{g}}\Theta\bigl(k_\floq{g}^2\bigr)\,,\\
\label{eqn:tauiny'-inclined}
\tau\ped{in}^{y'}&=0\,,
\end{align}
where we defined the coefficient $J^\floq{\ell_b}_{m_bm'}(\beta)$ as
\beq
J^\floq{\ell_b}_{m_bm'}(\beta)\equiv\frac{d_{m_b,m'+1}^\floq{\ell_b}(\beta)\sqrt{\ell_b(\ell_b+1)-m'(m'+1)}+d_{m_b,m'-1}^\floq{\ell_b}(\beta)\sqrt{\ell_b(\ell_b+1)-m'(m'-1)}}2\,.
\eeq
Finally, the contributions of the continuum and of the bound states can be added to get the total ionization torque: 
\beq
\tau\ped{ion}^{x'}=\tau\ped{out}^{x'}+\tau\ped{in}^{x'},\qquad\tau\ped{ion}^{y'}=0.
\label{eqn:tauion}
\eeq
To obtain the components of the torque in the $x$-$y$-$z$ frame, we simply need to apply a backwards rotation:
\beq
\label{eqn:rotation-tau}
\tau\ped{ion}^z=\tau\ped{ion}^{z'}\cos\beta-\tau\ped{ion}^{x'}\sin\beta\,,\qquad\tau\ped{ion}^x=\tau\ped{ion}^{x'}\cos\beta+\tau\ped{ion}^{z'}\sin\beta\,,\qquad\tau\ped{ion}^y=\tau\ped{ion}^{y'}=0\,.
\eeq
Note that, because $P\ped{ion}=\tau\ped{ion}^{z'}\Omega$, only one of the components of the torque is actually independent of $P\ped{ion}$. This is a direct consequence of having assumed a circular orbit for the binary.\footnote{Suppose that a force $\vec F$ acts on the companion, and let $\braket{\vec\tau}=\braket{\vec r\times\vec F}$ be the average torque over one orbit. Then, the average dissipated power is $\braket{P}=\braket{\vec v\cdot\vec F}=\braket{(\vec\Omega\times\vec r)\cdot\vec F}=\braket{(\vec r\times \vec F)\cdot\vec \Omega}=\braket{\vec\tau\cdot\vec\Omega}=\braket{\vec\tau}\cdot\vec\Omega$. This relation is identically satisfied by equations (\ref{eqn:pion-inclined}), (\ref{eqn:tauz'-inclined}) and (\ref{eqn:rotation-tau}). On the other hand, if we had continued with the approach outlined in (\ref{eqn:rotated-Y}), and neglected the change in the occupancy of the states with $m'\ne m_b$, we would have found a result that violates this identity, and that is therefore inconsistent.}

\subsection{Numerical Evaluation}

\begin{figure}
\centering
\includegraphics{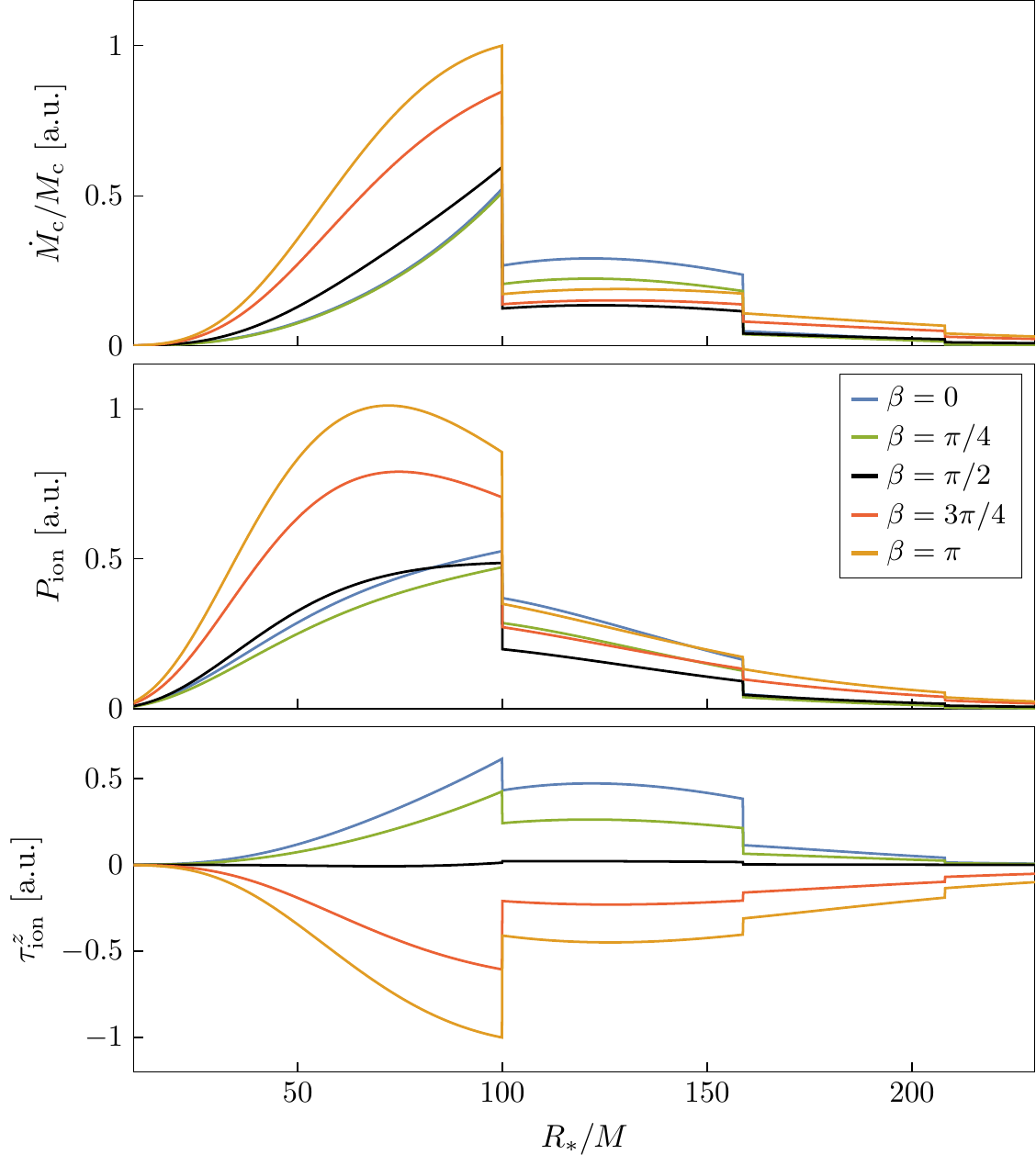}
\caption{Instantaneous ionization rate (\emph{top}), power (\emph{middle}) and torque along $z$ (\emph{bottom}) for a cloud in the $\ket{211}$ state, as function of the binary separation, for different values of the orbital inclination $\beta$. The $y$ axes are reported in arbitrary units (a.u.), while the $x$ axis has been normalized assuming $\alpha=0.2$.}
\label{fig:ionization_inclined_211}
\end{figure}

Expressions (\ref{eqn:rate-inclined}), (\ref{eqn:pion-inclined}), (\ref{eqn:tauz'-inclined}), (\ref{eqn:tauoutx'-inclined}) and (\ref{eqn:tauinx'-inclined}) can be evaluated numerically. In Figure~\ref{fig:ionization_inclined_211}, we show the ionization rate, power and $z$-component of the torque as function of the binary separation $R_*$, for selected values of the orbital inclination and a cloud in the $\ket{211}$ state. The same quantities for $\ket{322}$ are shown in Appendix~\ref{sec:ion322}.

\vskip 4pt
Varying $\beta$, each curves goes continuously from the equatorial co-rotating ($\beta=0$) to the equatorial counter-rotating ($\beta=\pi$) result of \cite{Baumann:2021fkf} (corrected with the $\ell_*=1$  dipole term \cite{wip_with_Rodrigo}). Rather than interpolating monotonically between the two limits, however, the curve is generally seen to first \emph{decrease} in amplitude, reaching a minimum for some intermediate value of the inclination (which varies depending on $R_*$), then increase again. This behaviour has an easy qualitative interpretation: the angular structure of the $\ket{211}$ state is such that the cloud has its highest density on the equatorial plane. When the binary's orbit is inclined, the companion does not stay in this high density region all the time, instead it moves out of it during parts of its orbit. According to the interpretation from Section~\ref{sec:dynamical-friction}, ionization is thus expected to be less efficient, because the companion encounters, on average, a lower local scalar density.

\subsection{Evolution of Inclination}
\label{sec:evolution-beta}

In the same spirit as Section~\ref{sec:evolution-e}, we can now study the backreaction of ionization on inclined orbits, in a simplified setup where self gravity and mass loss of the cloud, as well as accretion on the companion, are neglected.\footnote{If we wanted to track the mass loss of the cloud, an extra complication would be present. While (\ref{eqn:rate-inclined}) gives the \emph{total} ionization rate, the individual states $\ket{n_b\ell_bm'}$ are each ionized at a different rate. The cloud is then forced to go to a mixed state, and we would need to track the occupancies of all the $2\ell_b+1$ quasi-degenerate states. As some of them are not superradiant, it may be necessary to include their decay in the evolution too.} The energy balance equation reads, once again,
\beq
\frac{qM^2}{2R_*^2}\frac{\dd R_*}{\dd t}=-P\ped{ion}-P_\slab{gw}\,.
\label{eqn:energy-balance-inclined}
\eeq
Because we are considering circular orbits, equation (\ref{eqn:energy-balance-inclined}) is equivalent to the balance of angular momentum along the $z'$ axis. Instead, the other two components of the torque give new information. First of all, from (\ref{eqn:tauion}) and (\ref{eqn:rotation-tau}) we see that the torque lies in the $x$-$z$ plane, as its component along the $y$ axis vanishes identically. The orbital angular momentum also has a vanishing $y$ component, which will thus remain zero during the evolution of the system. In other words, we draw the conclusion that ionization induces \emph{no precession} of the orbital plane, and the orbit's axis will only rotate in the $x$-$z$ plane.

\vskip 4pt
This rotation, quantified by the evolution of the inclination angle $\beta$, is determined by the $x'$ component of the equation,
\beq
qM\sqrt{\frac{MR_*}{1+q}}\,\frac{\dd\beta}{\dd t}=-\tau\ped{ion}^{x'}\,.
\label{eqn:dbetadt}
\eeq
To understand the magnitude of the evolution of inclination, it is convenient to combine (\ref{eqn:energy-balance-inclined}) and (\ref{eqn:dbetadt}) as
\beq
\frac{\dd\beta}{\dd R_*}=\frac{\sqrt{(1+q)M}\,\tau\ped{ion}^{x'}}{2R^{5/2}(P\ped{ion}+P_\slab{gw})}\,,
\label{eqn:dbetadR}
\eeq
which allows us to compute $\beta$ as a function of $R_*$. This defines a ``trajectory'' in the $(R_*,\beta)$ plane that the binary follows through its evolution.

\begin{figure}
\centering
\includegraphics{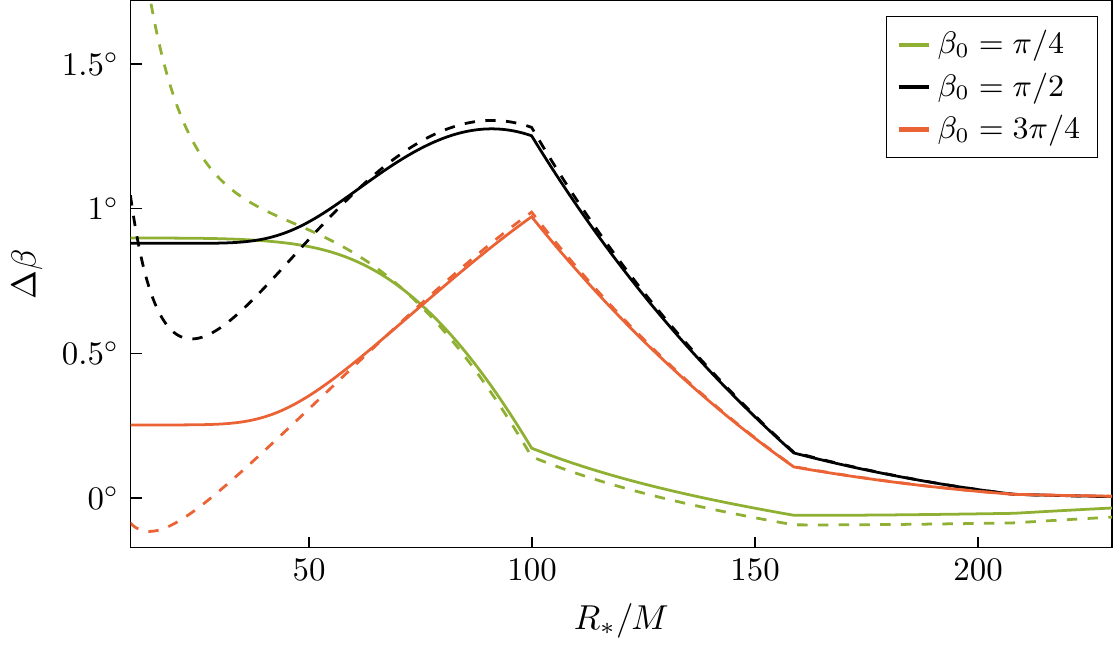}
\caption{Variation of the inclination angle $\Delta\beta\equiv\beta-\beta_0$, as function of the orbital separation $R_*$, for different values of the initial inclination $\beta_0$. The curves represent the evolution of $\Delta\beta$, from right to left, over the course of an inspiral. Solid lines are obtained by direct integration of (\ref{eqn:dbetadR}), with parameters $\alpha=0.2$, $q=10^{-3}$, $M\ped{c}=0.01M$ and a cloud in the $\ket{211}$ state. Dashed lines, instead, are computed by neglecting $P_\slab{gw}$ in (\ref{eqn:dbetadR}). In all cases, the inclination angle remains almost constant throughout the inspiral, with $\Delta\beta$ being at most of order 1 degree. We do not show trajectories with values of $\beta_0$ closer to 0 (co-rotating) of $\pi$ (counter-rotating), as the variation $\Delta\beta$ is even more limited in those cases.}
\label{fig:delta_beta_evolution}
\end{figure}

\vskip 4pt
As a general result, we find that the variation of the inclination angle $\beta$ is always very limited: over the course of a full inspiral, $\beta$ changes by at most a few degrees. It is useful to first consider the limit where ionization dominates the inspiral, thus neglecting $P_\slab{gw}$ in (\ref{eqn:dbetadR}). In this case, the  trajectory $\beta(R_*)$ only depends on the initial value $\beta_0\equiv\beta(R_*\to\infty)$, as well as on the state of the cloud. We show a few selected examples as dashed lines in Figure~\ref{fig:delta_beta_evolution}, where, for various choices of $\beta_0$, the total variation $\Delta\beta\equiv\beta-\beta_0$ is manifestly confined within a few degrees. When $P_\slab{gw}$ is included in (\ref{eqn:dbetadR}), the variation of $\beta$ is further limited: this case is shown with solid lines in Figure~\ref{fig:delta_beta_evolution}. This means that, as a simplifying approximation, the inclination angle $\beta$ can be treated as a fixed parameter in the evolution of the binary system.

\vskip 4pt
We conclude that, overall, ionization acts on inlined orbits in a simple way. The ionization rate (\ref{eqn:rate-inclined}) and power (\ref{eqn:pion-inclined}) need to be calculated for the specific value of the orbital inclination $\beta$ considered (see also Figure~\ref{fig:ionization_inclined_211}). The orbital plane, however, may be assumed to stay approximately fixed over time: the off-axis component of the torque induce no precession, and very little change in the value of $\beta$ over the course of an inspiral.

\section{Conclusions}
\label{sec:conclusions}

With the birth and development of GW astronomy, the scientific community became greatly interested in its potential as a tool for fundamental physics. One such application is the search for ultralight bosons produced by black hole superradiance, in particular the impact of a superradiant cloud on the dynamics of an inspiralling binary and the ensuing waveform. In the past few years, the phenomenology of these systems has been studied in detail, unveiling two distinct kinds of cloud-binary interaction: (1) resonant phenomena \cite{Zhang:2018kib,Baumann:2018vus,Baumann:2019ztm}, which occur at specific orbital frequencies, and (2) friction effects \cite{Zhang:2019eid,Baumann:2021fkf,Baumann:2022pkl}, which act continuously on the binary. In this paper, we study the latter, extending all previous studies in the direction of achieving a complete and coherent understanding of the evolution of the system.

\vskip 4pt
The process of binary formation via dynamical capture is altered by the cloud, which facilitates the formation of bound systems by opening up a new channel for energy dissipation. We demonstrate that this process is mediated by transitions to both bound and unbound states, with the latter generally giving the dominant contribution. As a consequence, the cross section for dynamical capture is increased up to a factor of $\mathcal O(10)-\mathcal O(100)$, compared to the case where only dissipation through GWs is taken into account. The capture and merger rates are correspondingly increased.

\vskip 4pt
Once the binary is formed, it proceeds to inspiral. Its time-varying gravitational perturbation causes the cloud to be ionized, taking energy from the system, whose orbital separation shrinks faster than expected in vacuum. We demonstrate quantitatively that this process should be understood as dynamical friction for the case of a gravitational atom. We then switch on orbital eccentricity and inclination in order to understand the effects of ionization in generic scenarios. We show that, on eccentric orbits, the energy losses kick in at larger separations (i.e.\ at an earlier stage of the inspiral) compared to the quasi-circular case. We also prove that the backreaction of ionization strongly circularizes the binary, except at separations that are so small that GWs are likely to dominate. Consequently, we conclude that the approximation of quasi-circular orbits, for the late stage of the inspiral, is justified. Finally, we repeate a similar exercise with orbital inclination. We find that, although inclination needs to be taken into account to compute the effect of ionization accurately, its value barely changes throughout the inspiral. We therefore conclude that treating the inclination angle as a fixed parameter is a well-justified approximation.

\vskip 4pt
Although our analysis is more general than previous studies, we still make a number of simplifying assumptions. Most notably, our results are entirely non-relativistic, both concerning the cloud's model and the orbital evolution. Relativistic corrections will be needed to properly describe the final stages of the inspiral, where, although the relative impact of the cloud fades away, the GW signal is louder and thus has the best chances of being detected. Moreover, for the sake of simplicity, we solved for the evolution of the orbital parameters in a simplified setup where the following effects were neglected: accretion of the cloud on the companion (if it is a BH) \cite{Unruh:1976fm,Benone:2019all,Baumann:2021fkf}, mass loss of the cloud due to ionization \cite{Baumann:2021fkf}, oscillatory transients occurring at the ionization discontinuities \cite{Baumann:2021fkf} and the backreaction of the cloud on the geometry \cite{Ferreira:2017pth,Hannuksela:2018izj}. The inclusion of all of these effects is straightforward and can be done in a more detailed numerical study.

\vskip 4pt
As anticipated, friction effects are only one part of the puzzle, the other being orbital resonances. Together, they shape the evolution of the binary, whose history becomes intimately tied to that of the cloud. While in this work we mostly showed results for the fastest growing mode $\ket{211}$, in reality the state of the cloud changes over time depending on the resonant transitions it encounters. Having completed this study, we plan to consistently explore the interplay between the two kinds of interactions in an upcoming work, drawing a complete and coherent picture of the dynamical evolution of the system.

\paragraph{Acknowledgements}

We are grateful to Daniel Baumann for discussions and detailed feedback on the manuscript, and to Vitor Cardoso, Rodrigo Vicente and John Stout for discussions and comments. TS is supported by the VILLUM Foundation (grant no.~VIL37766), the Danish Research Foundation (grant no.~DNRF162), and the European Union’s H2020 ERC Advanced Grant ``Black holes: gravitational engines of discovery'' grant agreement no.~Gravitas–101052587.

\newpage
\appendix
\addtocontents{toc}{\protect\vskip24pt}

\section{Binary Eccentrification}
\label{sec:appA}

\begin{figure}
\centering
\includegraphics{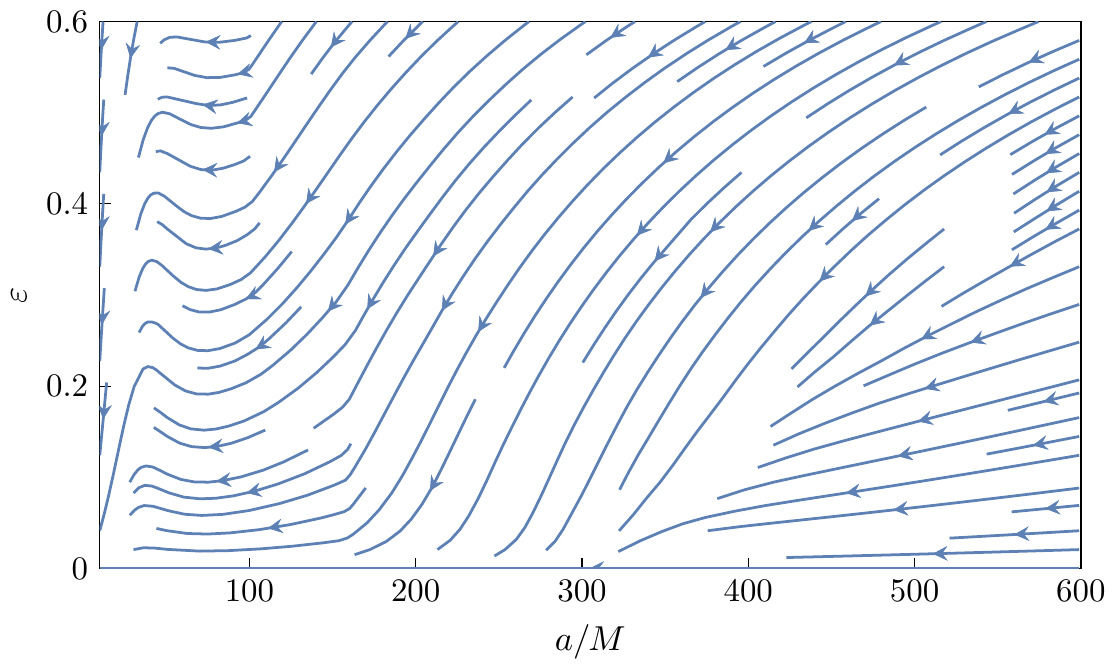}
\caption{Evolution of the eccentricity, including the effects of both ionization and GWs. The values of the parameters are the same as in Figure~\ref{fig:streamplot_eccentricity_unified}, except for the mass of the cloud, which has been increased from $M\ped{c}=0.01$ to $M\ped{c}=0.1M$.}
\label{fig:streamplot_eccentricity_with_GW_mass_10}
\end{figure}

As described in Section~\ref{sec:evolution-e}, ionization alone can eccentrify the binary if its semi-major axis is comparable to the size of the cloud. Though, the circularizing effect of GW emission generally washes that out. This conclusion, however, depends on the chosen parameters. In Figure~\ref{fig:streamplot_eccentricity_with_GW_mass_10}, we show an example where a small region of eccentrification survives. One can achieve that by simply increasing the mass of the cloud: this way, $P\ped{ion}$ is correspondingly increased, thereby shifting the tipping point where GWs take over to smaller values of $a$.

\section{Ionization on a Different State}
\label{sec:ion322}

\begin{figure}
\centering
\includegraphics{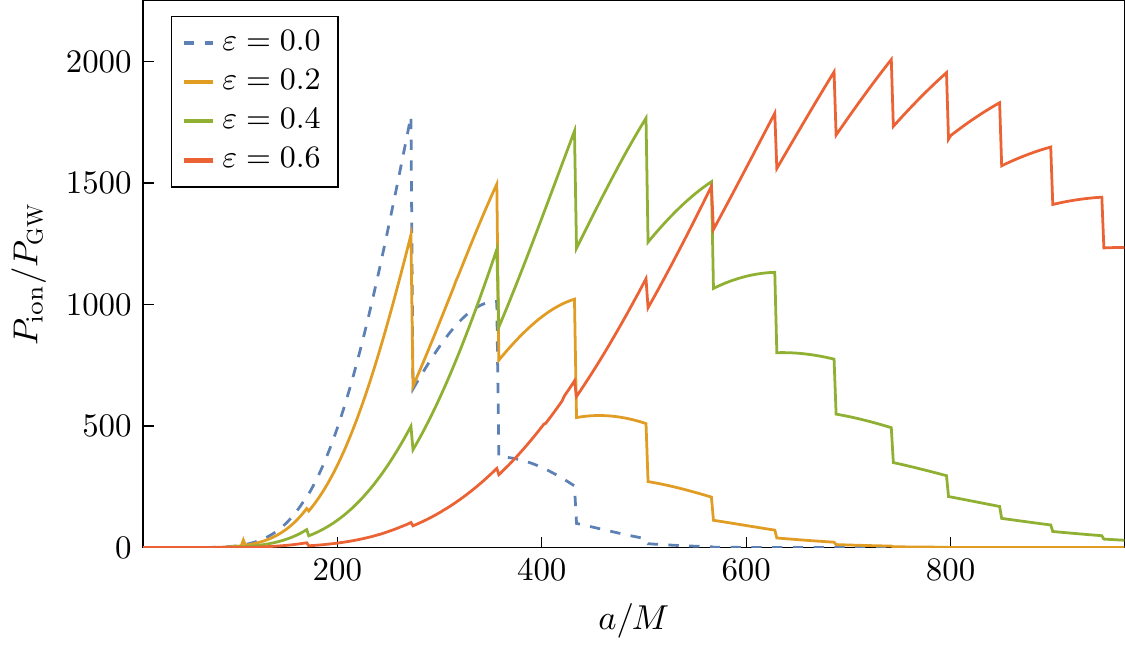}
\caption{Ionization power for different values of the eccentricity $\varepsilon$, for a cloud in the $\ket{322}$ state and a co-rotating equatorial orbit. The values of the other parameters are the same as in Figure~\ref{fig:P_ion_P_GW_eccentricity_211}.}
\label{fig:P_ion_P_GW_eccentricity_322}
\end{figure}

\begin{figure}
\centering
\includegraphics{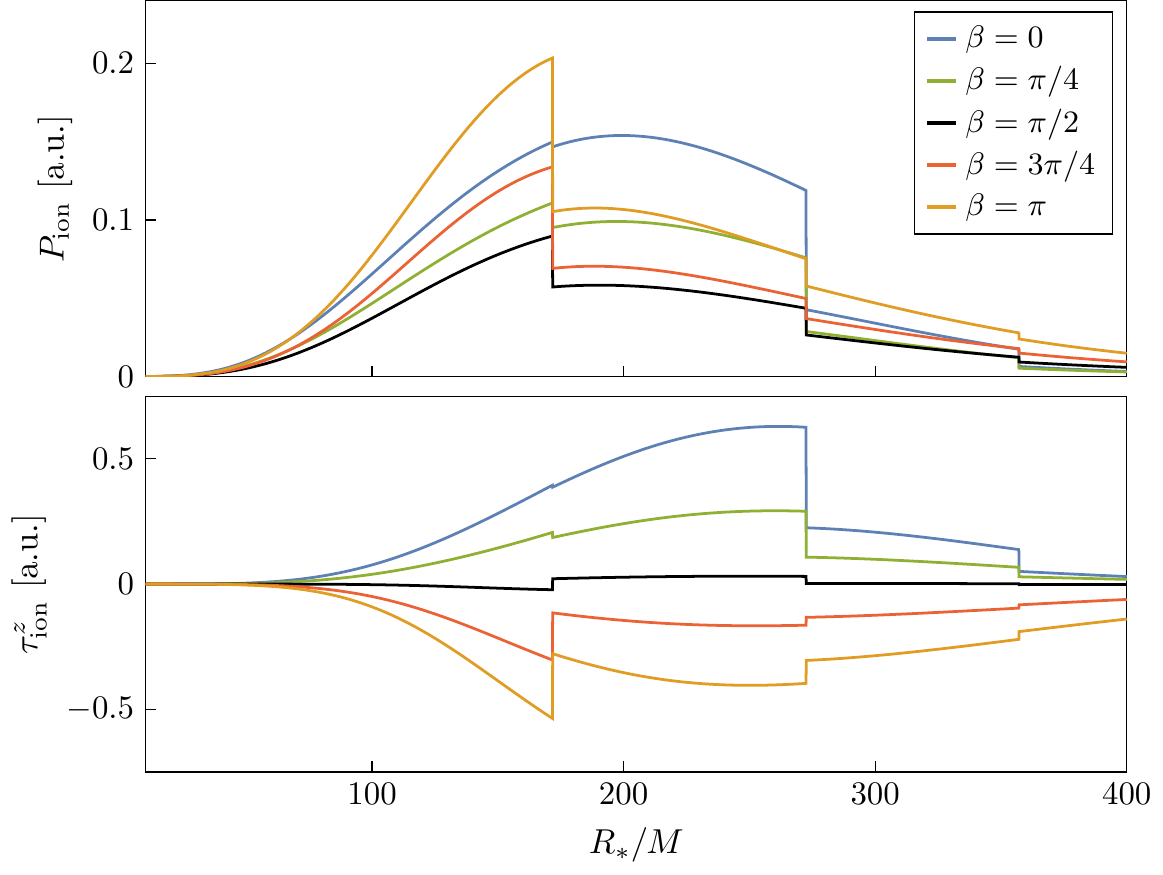}
\caption{Ionization power (\emph{top}) and torque along $z$ (\emph{bottom}) for a cloud in the $\ket{322}$ state, for different values of the inclination $\beta$. The values of the other parameters are the same as in Figure~\ref{fig:ionization_inclined_211}. The normalization of the curves is also the same as in Figure~\ref{fig:ionization_inclined_211}, so that the amplitudes can be directly compared between the two cases.}
\label{fig:ionization_inclined_322}
\end{figure}

In the main text, we show results for the ionization power and torque assuming a $\ket{211}$ state. In general, the cloud is expected to be in a state that not only depends on the initial mass and spin of the black hole, but also on the resonances the binary encounters during its evolution. While we postpone the study of this process to a future work, for completeness we show here the results assuming that the cloud is in a $\ket{322}$ state, which is generally the second fastest growing mode. Figures~\ref{fig:P_ion_P_GW_eccentricity_322} and \ref{fig:ionization_inclined_322} show the ionization power for eccentric and inclined orbits, respectively. Generally speaking, when $n_b$ is increased, the cloud moves farther away from the central BH and, as a consequence, it becomes more dilute. This reduces the ionization power. However, $P_\slab{gw}$ decreases even faster with increasing $R_*$, and thus the relative impact of ionization actually increases when the cloud is in an excited state.

\newpage

\clearpage

\phantomsection

\addtocontents{toc}{\protect\vskip24pt}
\addcontentsline{toc}{section}{References}

\makeatletter

\interlinepenalty=10000

{%\linespread{1.069}
\bibliographystyle{utphys}
\bibliography{main}
}
\makeatother

\end{document}